\documentclass[twocolumn]{aastex63}
\DeclareMathAlphabet\mathbfcal{OMS}{cmsy}{b}{n}
\usepackage{graphicx,amsmath}

\accepted{\today}
\submitjournal{ApJ}

\shorttitle{WISE 1828}
\shortauthors{Cushing et al.}

\newcommand\teff{\mbox{$T_\mathrm{eff}$}}
\newcommand\fsed{\mbox{$f_\mathrm{sed}$}}
\newcommand\logg{\mbox{$\log g$}}

\newcommand\thetaatm{\mbox{$\boldsymbol \theta_\mathrm{atm}$}}

\newcommand{\thetaatmx}[1]{\mbox{$\boldsymbol \theta_{\mathrm{atm},#1}$}}

\newcommand{\UGPSzeroseventwotwo}{UGPS 0722$-$05}

\newcommand{\WISEeighteentwentyeightFull}{\mbox{WISEP J182831.08$+$265037.8}}
\newcommand{\WISEeighteentwentyeight}{\mbox{WISE 1828$+$2650}}

\newcommand{\WISEzeroeightfivefive}{\mbox{WISE 0855$-$0714}}

\newcommand{\WISE}{\textit{WISE}}

\begin{document}

\title{An Improved Near-Infrared Spectrum of the Archetype Y Dwarf WISEP J182831.08$+$265037.8}

\correspondingauthor{Michael Cushing}
\email{michael.cushing@utoledo.edu}

\author[0000-0001-7780-3352]{Michael C. Cushing}
\affiliation{Ritter Astrophysical Research Center, Department of Physics and Astronomy, University of Toledo, 2801 W. Bancroft St., Toledo, OH 43606, USA}

\author[0000-0002-6294-5937]{Adam C. Schneider}
\affiliation{US Naval Observatory, Flagstaff Station, P.O. Box 1149, Flagstaff, AZ 86002, USA}

\author[0000-0003-4269-260X]{J. Davy Kirkpatrick}
\affiliation{IPAC, Mail Code 100-22, Caltech, 1200 E. California Blvd., Pasadena, CA 91125, USA}

\author[0000-0002-4404-0456]{Caroline V. Morley}
\affiliation{Department of Astronomy, University of Texas at Austin, Austin, TX, USA}

\author[0000-0002-5251-2943]{Mark Marley}
\affiliation{NASA Ames Research Center, NS 245-3, Moffett Field, CA, 94035}

\author{Christopher R. Gelino}
\affiliation{IPAC, Mail Code 100-22, Caltech, 1200 E. California Blvd., Pasadena, CA 91125, USA}

\author[0000-0001-7875-6391]{Gregory N. Mace}
\affiliation{McDonald Observatory and Department of Astronomy, University of Texas at Austin, 2515 Speedway, Stop C1400, Austin, TX 78712-1205, USA}

\author[0000-0001-5058-1593]{Edward L. Wright}
\affiliation{Department of Physics and Astronomy, UCLA, 430 Portola Plaza, Box 951547, Los Angeles, CA 90095-1547, USA}

\author{Peter R. Eisenhardt}
\affiliation{Jet Propulsion Laboratory, California Institute of Technology, 4800 Oak Grove Dr., Pasadena, CA 91109, USA}

\author[0000-0001-8671-5901]{Michael F. Skrutskie}
\affiliation{Department of Astronomy, University of Virginia, Charlottesville, VA, 22904}

\author[0000-0003-0107-7803]{Kenneth A. Marsh}
\affiliation{IPAC, Mail Code 100-22, Caltech, 1200 E. California Blvd., Pasadena, CA 91125, USA}



\begin{abstract}

  We present a \textit{Hubble Space Telescope}/Wide-Field Camera 3 near infrared spectrum of the archetype Y dwarf WISEP 182831.08$+$265037.8.  The spectrum covers the 0.9--1.7 $\mu$m wavelength range at a resolving power of $\lambda/\Delta \lambda \approx 180$ and is a significant improvement over the previously published spectrum because it covers a broader wavelength range and is uncontaminated by light from a background star.  The spectrum is unique for a cool brown dwarf in that the flux peaks in the $Y$, $J$, and $H$ band are of near equal intensity in units of $f_\lambda$.  We fail to detect any absorption bands of NH$_3$ in the spectrum, in contrast to the predictions of chemical equilibrium models, but tentatively identify CH$_4$ as the carrier of an unknown absorption feature centered at 1.015 $\mu$m.  Using previously published ground- and spaced-based photometry, and using a Rayleigh Jeans tail to account for flux emerging longward of 4.5 $\mu$m, we compute a bolometric luminosity of $\log (L_\mathrm{bol}/\mathcal{L}^N_\odot)=-6.50\pm0.02$ which is significantly lower than previously published results.  Finally, we compare the spectrum and photometry to two sets of atmospheric models and find that best overall match to the observed properties of \WISEeighteentwentyeight\ is a $\sim$1 Gyr old binary composed of two \teff$\sim$325 K, $\sim$5 $M_\textrm{Jup}$ brown dwarfs with subsolar [C/O] ratios.

\end{abstract}



\section{Introduction} \label{sec:intro}

The search for cool brown dwarfs with effective temperatures (\teff) less than 700 K (approximately a spectral type of T8) was one of the driving science goals of wide-area, red-optical and infrared surveys such as the United Kingdom Infrared Deep Sky Survey \citep[UKIDSS;][]{2007MNRAS.379.1599L}, the Canada France Brown Dwarf Survey \citep[CFBDS;][]{2008A&A...484..469D} and its counterpart in the near-infared CFBDSIR \citep{2010A&A...518A..39D}, and the Wide-field Infrared Survey Explorer \citep[\WISE;][]{2010AJ....140.1868W}.  With masses less than $\sim$20 $M_\mathrm{Jup}$ at the typical age of the ultracool dwarf field population of 3 Gyr \citep{2002AJ....124.1170D}, such cool brown dwarfs are most likely the least massive products of star formation. 

While such cool brown dwarfs have been discovered using each of these surveys \citep[e.g.,][]{2008MNRAS.391..320B,2008A&A...482..961D,2010A&A...518A..39D,2011ApJ...740..108L,2011ApJS..197...19K,2011ApJ...743...50C} the vast majority of brown dwarfs with effective temperatures less than 500 K have been discovered using \WISE\ data.  The near-infrared spectrum of one such brown dwarf, WISEP J182831.08$+$265037.8 (hereafter \WISEeighteentwentyeight), was distinct enough from that of the late-type T dwarfs that \citet{2011ApJ...743...50C} identified it has the archetype of the new Y spectral class.  They originally assigned it a spectral type of $>$Y0 but its type was later revised to $\geq$Y2 after the discovery of the Y1 dwarf WISE 035000.32$-$565830.2 by \citet{2012ApJ...753..156K}.  At the time of its discovery, \WISEeighteentwentyeight\ had the reddest near- to mid-infrared color of any brown dwarf known at $J$$-$$W2$$=9.29\pm0.35$ (mag).  Atmospheric models indicated such a color corresponded to an effective temperature of less than $\sim$300 K.

Once its parallax was measured \citep{2013ApJ...764..101B,2013Sci...341.1492D}, it became clear that \WISEeighteentwentyeight\ was overluminous in nearly all color magnitude and color spectral type diagrams \citep{2012ApJ...753..156K,2013ApJ...764..101B,2013ApJ...763..130L,2013Sci...341.1492D}.  Several reasons for this overluminosity were suggested including misclassification \citep{2012ApJ...753..156K}, a brightening in the spectral sequence due to an unknown physical mechanism \citep{2012ApJ...753..156K,2014ASSL..401..113C}, and  unresolved binarity \citep{2013ApJ...764..101B,2013ApJ...763..130L}.  The subsequent measurement of the parallaxes for the even redder and cooler brown dwarfs WD 0806$-$661B \citep{2011ApJ...730L...9L} and WISE J085510.83$-$071442.5 \citep[hereafter \WISEzeroeightfivefive,][]{2014ApJ...786L..18L} strongly suggest binarity is the cause of \WISEeighteentwentyeight's over luminosity because these two objects appear to fall in line with the other Y dwarfs in color magnitude diagrams \citep{2014ApJ...793L..16F,2014ApJ...796...39T,2015ApJ...799...37L,2016ApJ...823L..35S}.  Subsequent discoveries of brown dwarfs with similar effective temperatures like CWISEP J193518.59$-$154620.3 \citep{2019ApJ...881...17M}, CWISEP J144606.62$-$231717.8 \citep{2020ApJ...889...74M,2020ApJ...888L..19M} and WISEA J083011.95$+$283716.0 \citep{2020ApJ...895..145B}, have only strengthed the case for binarity but \WISEeighteentwentyeight\ remains stubbornly unresolved in both \textit{Hubble Space Telescope} (\textit{HST}) and ground-based adaptive optics imaging \citep{2011ApJ...743...50C,2013ApJ...764..101B}.  However given the poor quality of model atmosphere fits to the near- to mid-infrared spectral energy distribution of Y dwarfs \citep[e.g.,][]{2015ApJ...804...92S,2016AJ....152...78L,2017ApJ...842..118L}, missing or incomplete opacity sources and chemistry cannot be ruled out.

The spectrum of \WISEeighteentwentyeight\ presented by \citet{2011ApJ...743...50C} was obtained with the G141 grism in the Wide Field Camera 3 \citep[WFC3;][]{2008SPIE.7010E..43K} on-board the \textit{HST} and covered 1.1 to 1.7 $\mu$m at a resolving power of $R\equiv \lambda/\Delta \lambda \approx$ 130.  However because WFC3 is slitless, extreme care must be taken in order to avoid contamination by light from stars nearby on the sky.  At the time of the observations, we had yet to perfect our methodology for roll angle selection and so the resulting spectrum was contaminated by light from such a star.  In this paper, we present an improved, contamination-free \textit{HST}/WFC3 spectrum that covers an even larger wavelength range of 0.9--1.7 $\mu$m.

In \S2, we describe the acquisition and reduction of the spectrum.  In \S3, we discuss the characteristics of this unique spectrum in detail, revisit the spectral classification of \WISEeighteentwentyeight\, based on our new spectrum, and construct a spectral energy distribution covering 0.9--4.5 $\mu$m using previously published photometry.  \S4 presents our estimates for the bolometric luminosity and effective temperature of \WISEeighteentwentyeight.  The latter quantity was determined using both evolutionary models and two different sets of atmospheric models.  Finally in \S5, we discuss the model spectra fits to the data and the implications for the properties of \WISEeighteentwentyeight.

\section{Observations and Data Reduction}

\label{sec:obs}

\WISEeighteentwentyeight\ was observed with the WFC3 on-board \textit{HST} as a part of a Cycle 20 program to obtain grism spectroscopy of late-type T and Y dwarf candidates (GO-12970, PI=Cushing).  The WFC3 uses a 1024 $\times$ 1024 HgCdTe detector with a plate scale of 0$\farcs$13 pixel$^{-1}$ which results in a field of view of $123''\times 126''$.  \WISEeighteentwentyeight\ was observed with the G102 grism, which covers the 0.8$-$1.15 $\mu$m wavelength range at a resolving power of $R \equiv \Delta \lambda / \lambda$$\sim$210, and the G141 grism, which covers the 1.075$-$1.70 $\mu$m wavelength range at a resolving power of $R$$\sim$130.  Direct images through the F105W ($\lambda_\mathrm{pivot}$=1.0552 $\mu$m, $\Delta \lambda=0.265$ $\mu$m) or F125W ($\lambda_\mathrm{pivot}$=1.2486 $\mu$m, $\Delta \lambda =$ 0.2845 $\mu$m) filter were also obtained before the G102 and G141 grism images, respectively.  A log of the observations is given in Table \ref{tab:observations}.  Both the spectroscopic and photometric observations were reduced as described in \citet{2015ApJ...804...92S};  the resulting G102$+$G141 spectrum is shown in Figure 1 while the F105W and F125W magnitudes are given in Table \ref{tab:obs}.  There is a 2\% uncertainty in the absolute flux calibration of the spectrum \citep{2011wfc..rept....5K} and the signal-to-noise ratio (S/N) of the spectrum at the peaks of the $Y$, $J$, and $H$ bands (1.05, 1.27, and 1.6 $\mu$m), ranges from 15 to 20.  We also computed the F140W ($\lambda_\mathrm{pivot}$=1.3923$\mu$m, $\Delta \lambda=0.384$ $\mu$m) magnitude of \WISEeighteentwentyeight\ using the direct images obtained with the original G141 grism spectrum and that is also given in Table \ref{tab:obs}.  

\begin{deluxetable}{lcc}
\tablecolumns{3}
\tablewidth{0pc}
\tablecaption{\label{tab:observations}Log of \textit{HST}/WFC3 Observations}
\tablehead{
\colhead{UT Date} & 
\colhead{Grism/Filter} & 
\colhead{Total Integration Time (sec)}}

\startdata 
2013-Apr-22 & F105 & 509  \\
2013-Apr-22 & G102 & 7210 \\
2013-May-06 & F125 & 559  \\
2013-May-06 & G141 & 7210 \\
2013-May-08 & F105 & 509  \\
2013-May-08 & G102 & 7210 \\
2013-Jun-21 & F105 & 509  \\
2013-Jun-21 & G102 & 7210 \\
2013-Jun-25 & F105 & 509  \\
2013-Jun-25 & G102 & 7210 \\
2013-Aug-14 & F125 & 559  \\
2013-Aug-14 & G141 & 7210 \\
2013-Aug-17 & F105 & 509  \\
2013-Aug-17 & G102 & 7210  \\
\enddata 

\end{deluxetable}

\begin{deluxetable}{lcc}
\tablecolumns{3}
\tablecaption{\label{tab:obs}Summary of \label{tab:properties}\WISEeighteentwentyeight\ Photometry}
\tablehead{
\colhead{Parameter} & 
\colhead{Value} & 
\colhead{Reference}}

\startdata 
F105W  & 23.95$\pm$0.09 mag & this work \\
F125W  & 23.82$\pm$0.12 mag & this work \\
F140W  & 23.15$\pm$0.14 mag & this work \\
$z$(AB) & $>$24.46 mag & \citet{2013AA...550L...2L} \\
$Y$    & 23.20$\pm$0.17 mag & \citet{2013ApJ...763..130L} \\
$J$    & 23.48$\pm$0.23 mag &\citet{2013ApJ...763..130L} \\
$H$    & 22.45$\pm$0.08 mag & \citet{2013ApJ...764..101B} \\
$K$    & 23.48$\pm$0.36 mag & \citet{2013ApJ...763..130L} \\
$[$3.6$]$ & 16.915$\pm$0.020 & \citet{2019ApJS..240...19K} \\
$[$4.5$]$ & 14.321$\pm$0.020 & \citet{2019ApJS..240...19K} \\
AllWISE W1        & $>$18.248\tablenotemark{a} & \citet{2013wise.rept....1C} \\
AllWISE W2        & 14.353$\pm$0.045 & \citet{2013wise.rept....1C} \\
AllWISE W3        & 12.444$\pm$0.338\tablenotemark{b} &  \citet{2013wise.rept....1C} \\
AllWISE W4        & $>$8.505\tablenotemark{a}         &  \citet{2013wise.rept....1C} \\
CatWISE2020 W1        & 18.823$\pm$0.224\tablenotemark{c} & \citep{2021ApJS..253....8M} \\
CatWISE2020 W2        & 14.393$\pm$0.016 & \citep{2021ApJS..253....8M}  \\
\enddata 
\tablenotetext{a}{The flux measurement has S/N$<$2; the magnitude quoted is derived from the 95\% confidence flux upper limit.}
\tablenotetext{b}{S/N=3.2 with no clear source on the atlas tile.}
\tablenotetext{c}{S/N=2.9 but a faint source is visible and the proper motion of the source agrees with published results obtained with \textsl{Spitzer}.}

\end{deluxetable}

\begin{figure*} 
\centerline{\hbox{\includegraphics[width=6in,angle=0]{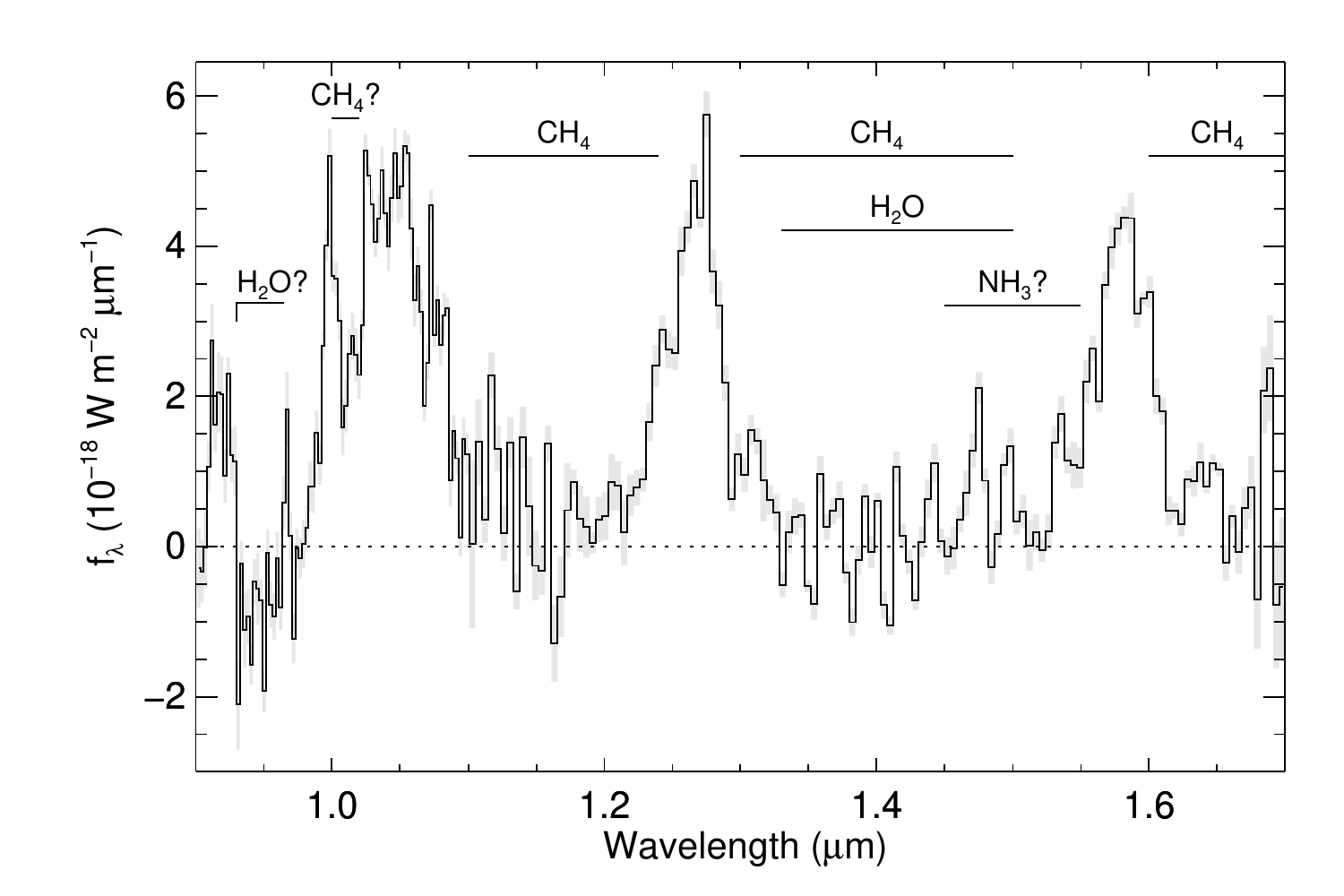}}}
\caption{\label{fig:W1828}New G102+G141 HST/WFC3 spectrum of \WISEeighteentwentyeight.  Prominent H$_2$O and CH$_4$ absorption bands are indicated along with potential absorption from the $\nu_2 + \nu_3$ NH$_3$ band from 1.45--1.55 $\mu$m and tentative 3($\nu_1,\nu_3$) H$_2$O band head at 0.925 $\mu$m.}
\end{figure*}

\section{The Spectrum}
\label{sec:thespectrum}
  
The spectrum exhibits deep absorption bands of CH$_4$ and H$_2$O typical of late-type T dwarf spectra \citep[e.g.,][]{2006ApJ...637.1067B} but the heights of the $Y$-, $J$-, and $H$-band peaks are approximately equal in units of $f_\lambda$.  This atypical characteristic prompted \citet{2011ApJ...743...50C} to identify \WISEeighteentwentyeight\, as the first Y dwarf.

We tentatively identify the 3($\nu_1,\nu_3$) band head of H$_2$O at 0.925 $\mu$m in the spectrum of \WISEeighteentwentyeight.  This band head is prominent in the spectra of the T dwarfs \citep{2003ApJ...594..510B,2011ApJS..197...19K} and has even been detected in the spectrum of the Y0 dwarf WISEPC J205628.90+145953.3 \citep{2013ApJ...763..130L}.  However given the low S/N at these wavelengths and the fact that the feature falls at the end of the spectrum, a higher S/N spectrum will be required in order to confirm the band head is real and carried by H$_2$O.

Ammonia is also a important opacity source in the atmospheres of cool brown dwarfs because it becomes the dominant equilibrium chemistry nitrogen-bearing species  (A(N$_2$)/A(NH$_3$) $<$ 1) at $T$ $\lesssim$ 700 K  \citep{2002Icar..155..393L}. The $\nu_2$ fundamental band is found in the spectra of T dwarfs at 10.5 $\mu$m \citep{2004ApJS..154..418R,2006ApJ...648..614C} but the overtone and combination bands in the near infrared (1.03, 1.21, 1.31, 1.51, 1.66, 1.98, and 2.26 $\mu$m) have defied clear detection because they are intrinsically weaker than the fundamental bands and often overlap in wavelength with strong H$_2$O and CH$_4$ overtone and combination bands.  

It has long been suggested that the emergence of these bands could trigger the creation of a new spectral class \citep{2003ApJ...596..587B,2007ApJ...667..537L,2008ASPC..384...85K} and indeed \citet{2011ApJ...743...50C} set the transition between the T and Y dwarfs where the \citet{2008A&A...482..961D}  NH$_3$-$H$ spectral index, which measures the strength of the NH$_3$ absorption band on the blue side of the $H$-band peak, indicated a significant increase in absorption.  \citet{2011AJ....142..169B} identified weak NH$_3$ features in a moderate resolution ($R=6000$) spectrum of the T9 dwarf \UGPSzeroseventwotwo\, across the entire near-infrared wavelength range but many of the identifications were later called into question by \citet{2012ApJ...750...74S}.  More recently, \citet{2019ApJ...877...24Z} measured the abundances of NH$_3$ in the atmospheres of 8 Y dwarfs through a retrieval analysis of the \citet{2015ApJ...804...92S} \textit{HST}/WFC3 spectra but no clear detection of NH$_3$ absorption features were reported.  

\begin{figure} 
\centerline{\hbox{\includegraphics[width=3.5in,angle=0]{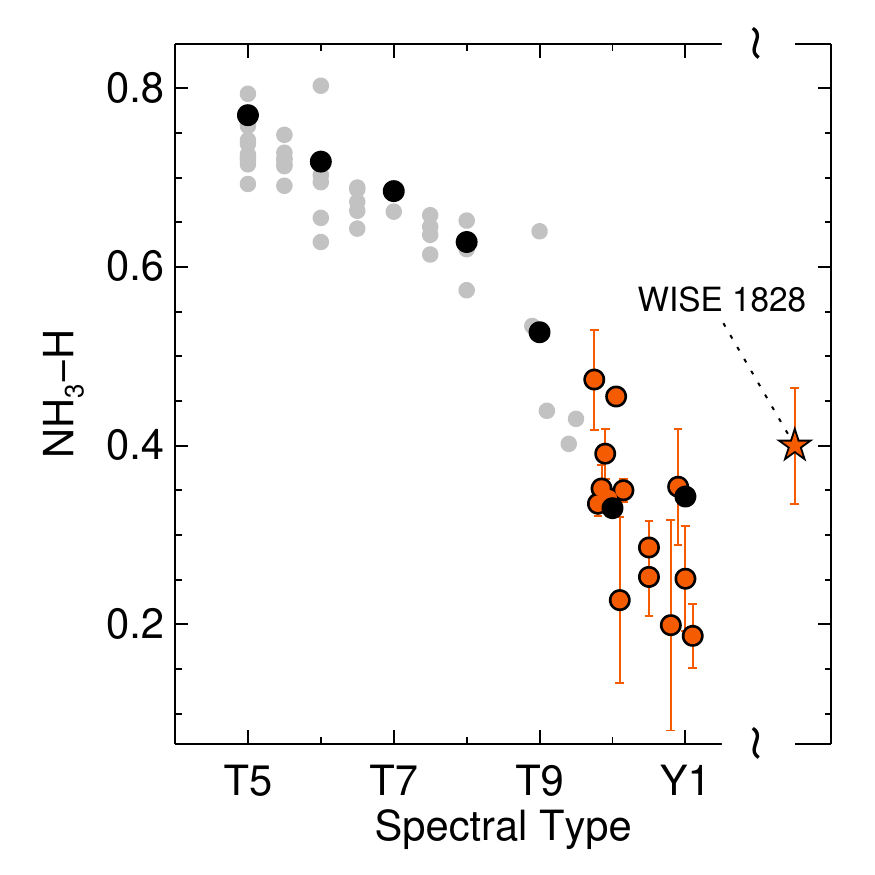}}}
\caption{\label{fig:nh3h} The \citet{2008A&A...482..961D} NH$_3$-$H$ spectral index as a function of spectral type.   The black points are for the T5--Y1 spectral standards.  The grey points were computed using SpeX/IRTF spectra from the SpeX Prism Library and the T dwarf spectra from \citet{2015ApJ...804...92S} while the red points were computed using the Y dwarf spectra from \citet{2015ApJ...804...92S} and from this work.}
\end{figure}

The peak absorption coefficients of the near-infrared NH$_3$ bands decrease steadily as one moves to shorter wavelengths \citep{2012ApJ...750...74S} which suggests that all else being equal, longer wavelength bands would be easier to detect.  While our spectrum does not extend to the $K$ band, a search for the 1.98 and 2.26 $\mu$m NH$_3$ bands would be difficult since the $K$-band flux of late-type T and Y dwarfs is heavily supressed due to strong CH$_4$ and collision-induced H$_2$ absorption.  The 1.66 $\mu$m band is centered beyond the wavelength limit of our spectrum but is also overwhelmed by CH$_4$ absorption and so would nevertheless be difficult to identify.  Visual inspection of the red side of the $H$-band peak shows no distinctive absorption feature due to the 1.51 $\mu$m NH$_3$ band, although this could be a result of the low S/N and/or low resolving power of the spectrum.  We have computed the NH$_3$-$H$ spectral index of \citet{2008A&A...482..961D} for \WISEeighteentwentyeight\ and a sample of T and Y dwarfs and the results are shown in Figure \ref{fig:nh3h}.  \WISEeighteentwentyeight\ deviates significantly from the trend set by the T and other Y dwarfs.  If the index is actually measuring NH$_3$ absorption, then the \WISEeighteentwentyeight\ value suggests that either NH$_3$ is absent from the spectrum, or the index is no longer sensitive to NH$_3$ at this spectral type.  The 1.21 and 1.31 $\mu$m NH$_3$ bands are difficult to identify because they overlap in wavelength with the CH$_4$ bands that give rise to the $J$-band emission peak and thus simply help to confine the emergent flux to an ever narrowing wavelength range.

In contrast, the 1.03 $\mu$m 2$\nu_1 + 2\nu_4$ band is centered near the peak of the $Y$-band which suggests it will be more easily identified.  However \citet{2015ApJ...804...92S} found no evidence of this feature in \textit{HST}/WFC3 spectra of other late-type T dwarfs and Y dwarfs.  The first  panel in the left column of Figure \ref{fig:opacity}  shows the spectrum of \WISEeighteentwentyeight\ centered on the $Y$-band peak.  There is a weak double-peaked absorption feature centered at 1.015 $\mu$m that at first appears to be  consistent with this NH$_3$ band. However also plotted in the second panel of the left column is the opacity cross-section spectrum of NH$_3$ at $T$=100 K and $P$=0.1 bar generated using the \citet{2011MNRAS.413.1828Y} NH$_3$ line list.  The position of the absorption feature in the spectrum of \WISEeighteentwentyeight\ clearly does not match the position of the NH$_3$ band.  Several possibilities present themselves:

\begin{figure*} 
\centerline{\hbox{\includegraphics[width=7in,angle=0]{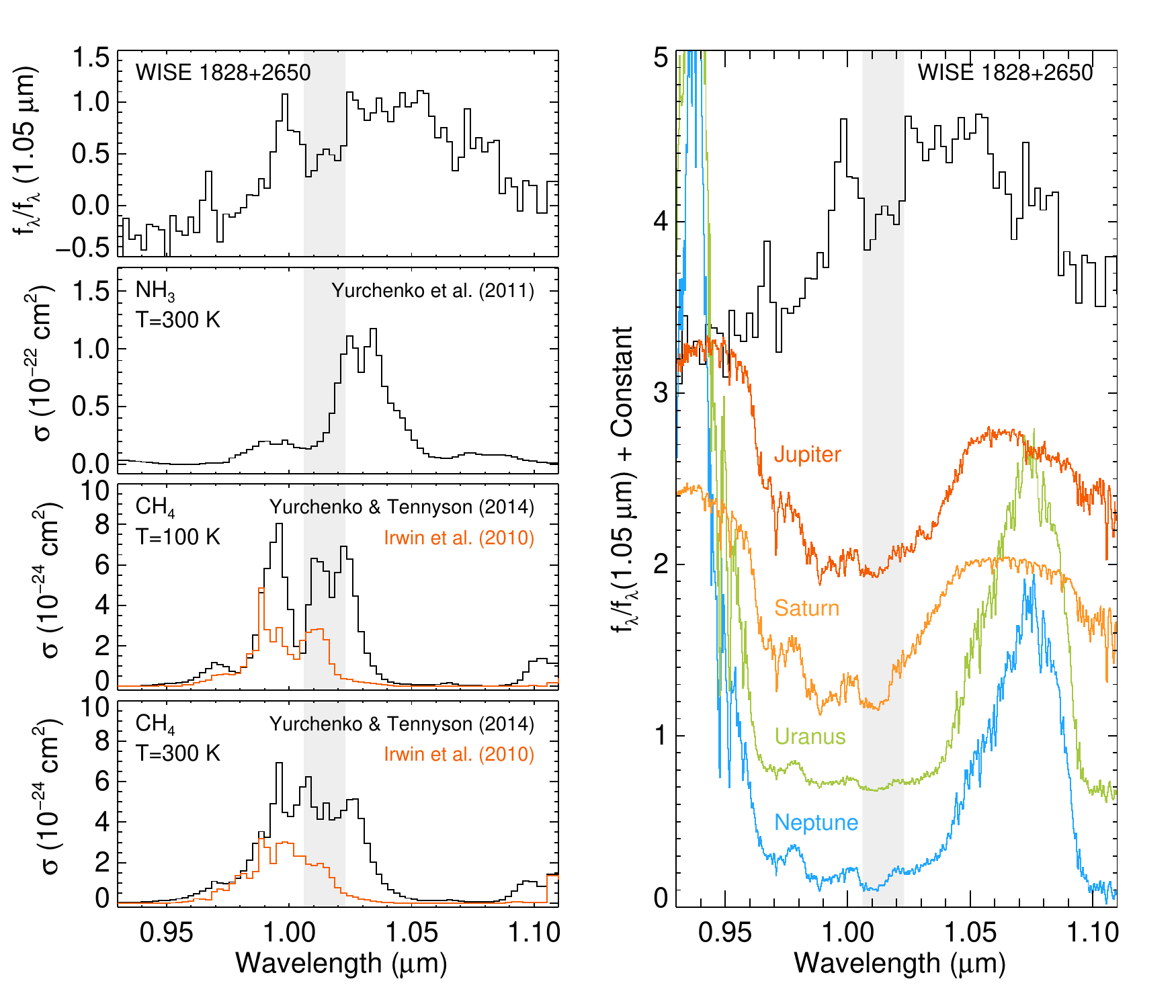}}}
\caption{\label{fig:opacity}Left: $Y$-band spectrum of \WISEeighteentwentyeight\ with the unknown absorption feature highlighted in light gray along with the cross-sections of CH$_4$ and NH$_3$ smoothed to the resolving power of $R$=210 and resampled onto the wavelength grid of the \WISEeighteentwentyeight\ spectrum.  Right:  $Y$-band spectrum of \WISEeighteentwentyeight\ with the unknown absorption feature highlighted in light gray  along with the reflected-light spectra of the gas giant planets from \citet{2009ApJS..185..289R}.  The giant planet spectra are at $R=2000$.}
\end{figure*}

\begin{enumerate}

\item \textit{The feature is real and is carried by NH$_3$ but the \citet{2011MNRAS.413.1828Y} line list is in error.}  Laboratory observations of NH$_3$ at similar temperatures and pressures by \citet{1969CoLPL...6..275C}, \citet{1999JQSRT..62..193I}, and \citet{2008Icar..196..612B} confirm the position of the band and its substructure therefore we eliminate this possibility. 

\item \textit{The feature is real and is carried by NH$_3$ but a wavelength calibration error in the \WISEeighteentwentyeight\ spectrum has shifted the feature blueward.}  We eliminate this as a possibility because other Y dwarf G102 spectra reduced by \citet{2015ApJ...804...92S} in the same manner do not appear discrepant with the spectra of other T/Y dwarfs.

\item \textit{The feature is not real and is an artifact of the data reduction.}  Such an ``absorption feature'' could arise in a low S/N spectrum if a cluster of bad or noisy pixels conspire to produce an emission-like feature at $\sim$1.0 $\mu$m thereby creating an apparent absorption feature at $\sim$1.03 $\mu$m.  In this case, the identification of the absorption feature would be a result of our a priori expectations of finding NH$_3$ absorption.  Five individual spectra are combined to produce the final spectrum and we can visually identify the absorption feature in four of the five spectra and so we eliminate this possibility.

\item \textit{The feature is real but its carrier is a molecule other than NH$_3$.} Since the wavelengths of the NH$_3$ absorption coefficients are correct, and we cannot identify an error in our data reduction, we conclude that the feature is probably real but is carriered by a molecule other than NH$_3$.  We explore this possibility in more detail presently.

\end{enumerate}

In a search for the carrier of this feature, we identified an absorption feature in the reflected-light spectra of Jupiter, Saturn, Uranus, and Neptune \citep{2009ApJS..185..289R} whose rough central wavelength and width matches that of the unidentified feature (see right panel of Figure \ref{fig:opacity}).   The presence of this band in all of the solar system giant planets strongly suggests the carrier is methane as ammonia is sequestered well below the clouds in Uranus and Neptune.  Indeed a comparison of the giant planet spectra to models generated with and without CH$_4$ opacity also indicate the feature is carried by CH$_4$ (P. Irwin, private communication).  The third and fourth panels of the left column of Figure \ref{fig:opacity} therefore show the cross-section spectrum of CH$_4$ at $T$=100 and 300 K ($P$=0.1 bar) from \citet{2014MNRAS.440.1649Y} and \citet{2010Icar..208..913I}; they have been smoothed to $R$=210 and resampled onto the wavelength grid of the \WISEeighteentwentyeight\ spectrum.  The $T$=100 K \citet{2014MNRAS.440.1649Y} spectrum provides the best match to structure of the absorption feature.  In particular, the double-peaked cross-section spectrum matches the structure of the absorption band and the ``emission-like'' feature at $\sim$1.0 $\mu$m is naturally explained by the lack of opacity at that wavelength.  However the \citeauthor{2014MNRAS.440.1649Y} band is shifted slightly redward of our feature which taken at face value weakens the case for the feature being carried by CH$_4$.  Interestingly, the position of the \citeauthor{2010Icar..208..913I} band does not match that of the \citeauthor{2014MNRAS.440.1649Y} band which suggests some uncertainty in the theoretical position of this band.  The apparent match between the low-temperature methane opacity structure and the observed feature is puzzling. For a cloudless atmosphere models indicate that the flux in this spectral region emerges from quite deep in the atmosphere, where the local temperature is around 800 K. If the carrier of the feature is methane, this may be an indication that water clouds, which form around 300 K, are limiting the depth of the visible atmosphere here. A complete analysis of this will be the subject of future work.

\subsection{Spectral Classification}

\WISEeighteentwentyeight\ was originally classified as $>$Y0 by \citet{2011ApJ...743...50C} because its 1.1$-$1.7 $\mu$m spectrum exhibited unique characteristics unseen in the spectra of late-type T dwarfs and Y0 dwarfs.  Its type was later revised to $\geq$Y2 after the discovery of the Y1 dwarf WISE 035000.32$-$565830.2 \citep{2012ApJ...753..156K}. With an improved spectrum in hand, we can return to the question of its spectral type.

\begin{figure} 
\centerline{\hbox{\includegraphics[width=3.5in,angle=0]{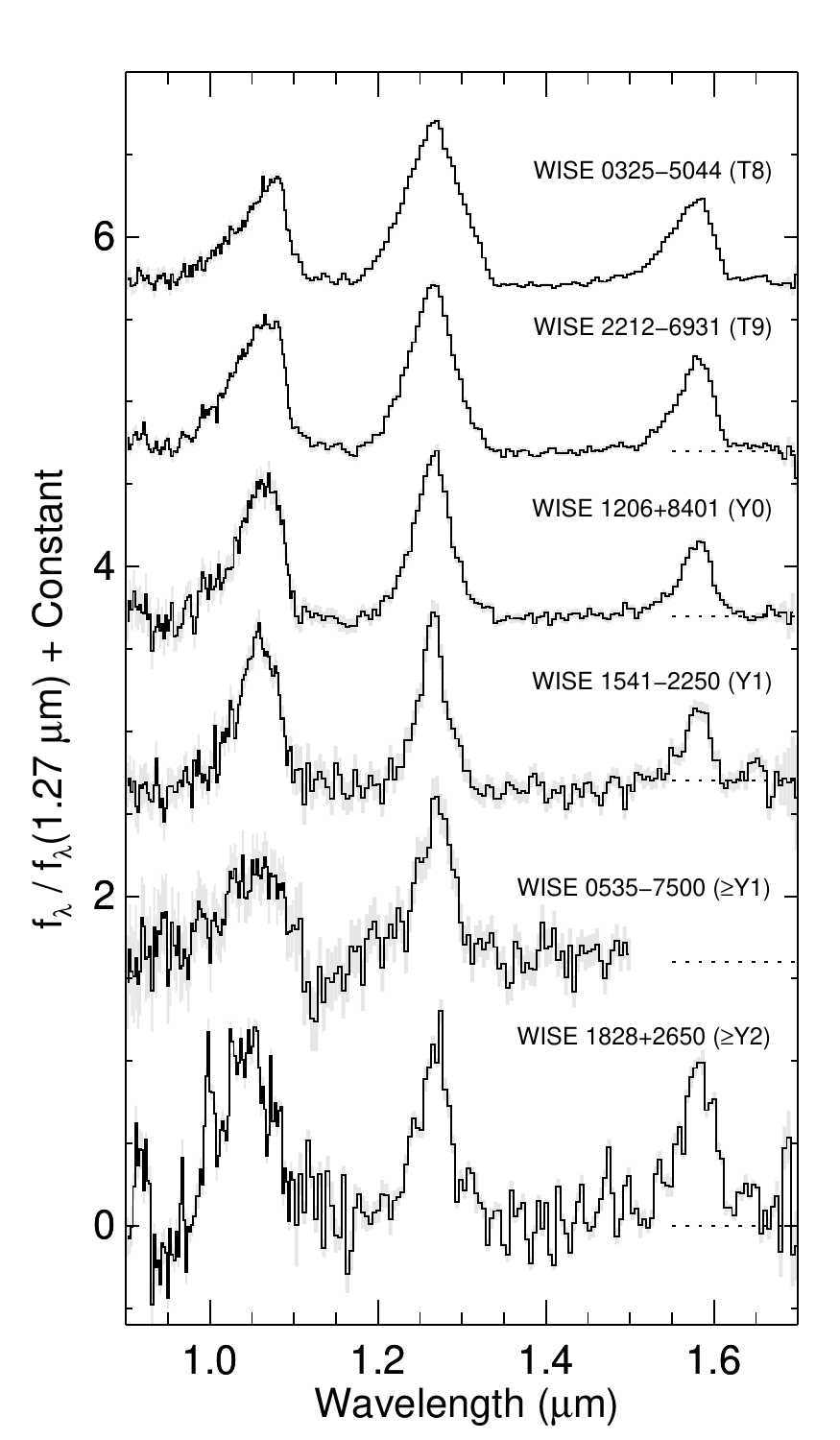}}}
\caption{\label{fig:sequence}0.9--1.7 $\mu$m spectral sequence at the T/Y boundary. The spectra of WISEA J032504.52$-$504403.0 (T8), WISEA J221216.27$-$693121.6 (T9), WISEA J120604.25$+$840110.5 (Y0), WISE J154151.65$-$225024.9 (Y1), and WISEA J053516.87$-$750024.6 ($\geq$Y1) are from \citet{2015ApJ...804...92S} while the spectrum of \WISEeighteentwentyeight\ is from this work.  The $H$-band spectrum of WISEA J053516.87$-$750024.6 is missing because it was contaminated by first-order light from a nearby star \citep{2015ApJ...804...92S}.  Uncertainties for each spectrum are shown as gray bars.}
\end{figure}

Figure \ref{fig:sequence} shows a sequence of late-type T and Y dwarf spectra all observed with the G102+G141 WFC3 grism combination \citep[][this work]{2015ApJ...804...92S}.  The new \WISEeighteentwentyeight\ spectrum is still clearly distinct from that of the T dwarfs and therefore the spectral type should remain a Y dwarf.  Trends in spectral morphology at the T/Y boundary noted by \citet{2012ApJ...753..156K} and \citet{2015ApJ...804...92S} continue with the addition of the \WISEeighteentwentyeight\ spectrum:  1) the $Y$-band peak becomes more symmetrical, 2) the $Y$-band peak wavelength becomes bluer, 3) the $H$-band peak becomes more symmetrical, and 4) with the exception of WISEA J053516.87$-$750024.6 (whose G102 spectrum may still be contaminated \citep{2015ApJ...804...92S}), the $Y$-, $J$-, $H$-band peaks evolve towards having equal intensities.  Interestingly, the tentative new CH$_4$ band appears weakly in the spectrum WISEA J053516.87$-$750024.6, although the S/N is quite low. Since the spectrum of \WISEeighteentwentyeight\ smoothly extends the Y dwarf spectral sequence beyond Y1, its spectral type should remain $\ge$Y2 until the near-infrared spectra of cooler objects like \WISEzeroeightfivefive\, are obtained.  

\subsection{The Spectral Energy Distribution}

\begin{figure*}
 \centerline{\hbox{\includegraphics[width=6in,angle=0]{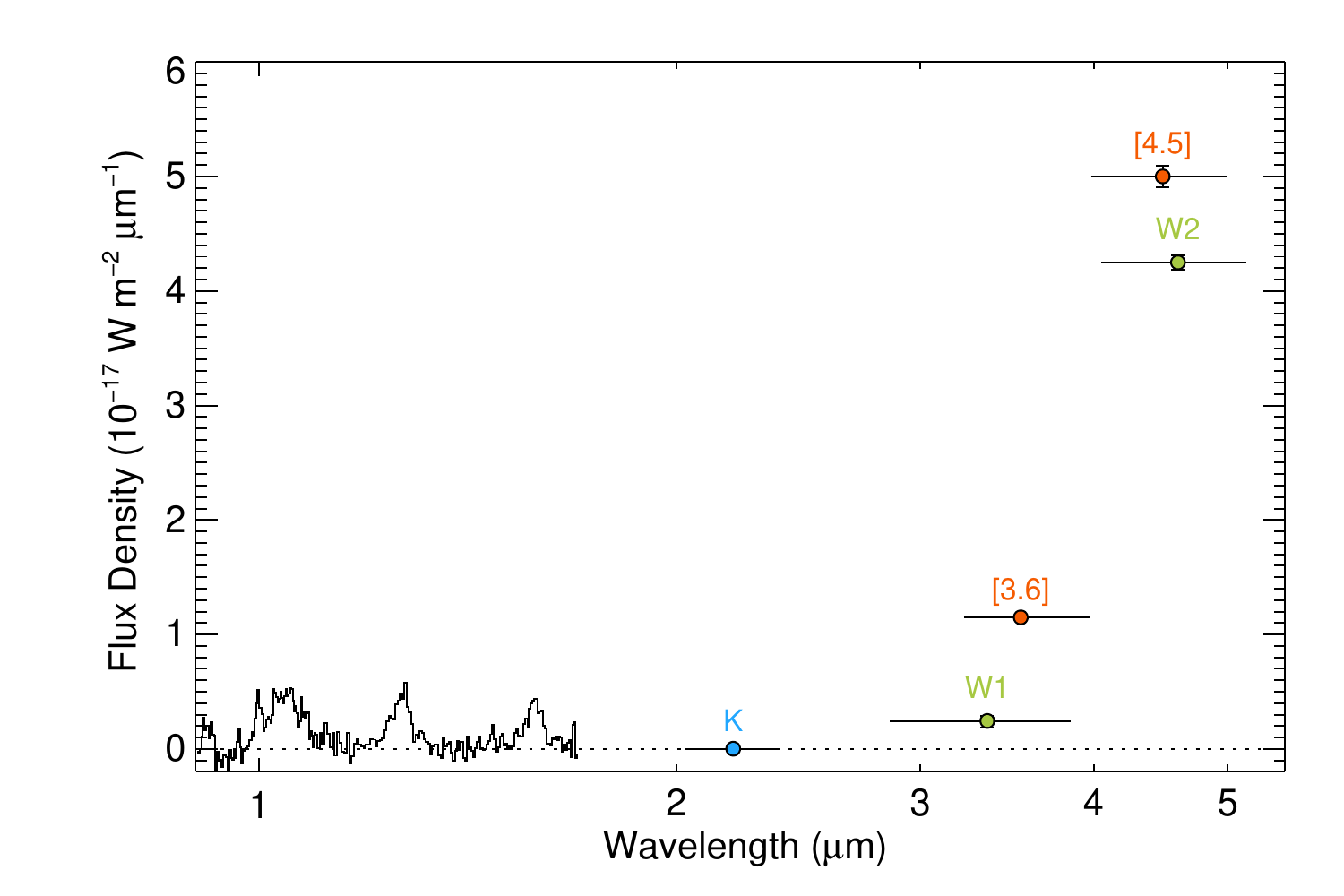}}} \caption{\label{fig:SED} Spectral energy distribution of \WISEeighteentwentyeight.  The spectrum (black) is from this work, while the $K$-band photometric point is from \citet{2013ApJ...763..130L}, the CatWISE2020 W1 and W2 photometry is from \citet{2021ApJS..253....8M}, and the \textit{Spitzer} [3.6] and [4.5] photometry is from \citet{2019ApJS..240...19K}.  The photometry is plotted at the nominal or isophotal filter wavelengths and the widths of the filters are denoted with horizontal lines.}
\end{figure*}

Figure \ref{fig:SED} shows the spectral energy distribution of \WISEeighteentwentyeight\ constructed by combining the new \textit{HST} spectrum and the MKO $K$-band \citep{2013ApJ...763..130L}, \textit{Spitzer} [3.6] and [4.5] \citep{2019ApJS..240...19K}, and CatWISE2020 $W1$ and $W2$ \citep{2021ApJS..253....8M}  photometric points.   The apparent magnitudes were converted to average flux densities and plotted at the wavelengths of $\lambda_\textrm{iso}^K$=2.198 $\mu$m, $\lambda_0^{[3.6]}$ = 3.544 $\mu$m, $\lambda_0^{[4.5]}$=4.487 $\mu$m, and $\lambda_\textrm{iso}^{W1}$ = 3.3526 $\mu$m, $\lambda_\textrm{iso}^{W2}$=4.602 $\mu$m using the zero points and wavelengths in \citet{2005PASP..117..421T}, \citet{2005PASP..117..978R}, and \citet{2011ApJ...735..112J}, for the MKO, \textit{Spitzer}, and \textit{WISE} photometry, respectively.  For clarity, we do not include photometric measurements at wavelengths that are sampled by the \textit{HST} spectrum.  The large amount of flux emerging at [4.5]/$W2$ is a result of the fact that this wavelength range is relatively free of stronger absorbers like CH$_4$, NH$_3$, and H$_2$O which allows the observer to see to deeper and thus hotter layers of the atmosphere were $T>T_\mathrm{eff}$.  

\section{Physical Properties}

\subsection{Bolometric Luminosity}
\label{sec:lbol}

A brown dwarf's bolometric luminosity is arguably the easiest fundamental parameter to measure since knowledge of only distance $d$ and bolometric flux $f_\mathrm{bol}$ is required.  The spectral energy distributions of hotter L and early- to mid-type T dwarfs peak at red-optical and near-infrared wavelengths where spectra are easily obtained from the ground.  Integration of a spectral energy distribution consisting of a red optical and near-infrared spectrum, an $L$-band photometric point and/or WISE W1 and W2 points, and a Rayleigh Jeans tail to account for (unobserved) emergent flux at longer wavelengths therefore provides a reasonably accurate estimate of $F_\mathrm{bol}$ \citep{2006ApJ...648..614C,2015ApJ...810..158F}.  However the shift in the peak of the Planck function to longer wavelengths for the cooler late-type T dwarfs and Y dwarfs complicates the calculation of their bolometric fluxes because more and more energy emerges at wavelengths that are either difficult or impossible to access with current instrumentation \citep{2009and..book..101M}.

We constructed a continuous spectral energy distribution of \WISEeighteentwentyeight\ as follows.  We first assumed that no flux emerges at wavelengths shortward of 0.90 $\mu$m and then extended the \textit{HST} spectrum by piecewise linear interpolation through the $K$-band, $W1$, [3.6], [4.5], and $W2$ photometric points.  \WISEeighteentwentyeight\ has W3=12.444$\pm$0.338 \citep[S/N=3.2,][]{2013wise.rept....1C} but visual inspection of the atlas tile shows no clear source.  We therefore do not use it or the W4 upper limit in the construction of the spectral energy distribution.  To account for the flux at wavelengths longer than $W2$, we extend a Rayleigh Jeans tail from the flux density point at 4.602 $\mu$m  (the isophotal wavelength of the $W2$ filter)  to $\lambda=\infty$.  Integrating over the spectral energy distribution, we measure a bolometric flux of $f_\mathrm{bol}$ = $(1.015\pm0.01) \times 10^{-16}$  W m$^{-2}$, which is the mean and standard deviation of 10,000 Monte Carlo realizations of $f_\mathrm{bol}$ that account for the 2\% uncertainty in the absolute flux calibration of the \textit{HST} spectrum and the uncertainties in the individual spectral and photometric points.  Using a relative parallax of $\pi=100.7\pm2.3$ mas \citep{2019ApJS..240...19K}, we find $\log (L_\mathrm{bol}/\mathcal{L}^N_\odot)=-6.50\pm0.02$, where $\mathcal{L}^N_\odot$ is the International Astronomical Union (IAU) nominal solar luminosity of $3.828 \times 10^{26}$ W  \citep{2015arXiv151007674M}. We urge caution when using the $f_\textrm{bol}$ and $L_\mathrm{bol}/\mathcal{L}^N_\odot$ measurements because they are likely dominated by systematic uncertainties due to the fact that the emergent spectra of cool brown dwarfs at wavelengths longward of $W2$ are not Planck functions \citep[e.g.,][]{2003ApJ...596..587B,2006ApJ...648..614C} and the Rayleigh-Jeans tail extension contributes $\sim$62\% of the total bolometric flux in units of $f_\lambda$.

\citet{2013Sci...341.1492D} computed the bolometric luminosity of \WISEeighteentwentyeight\ using near-infrared and \textit{Spitzer Space Telescope}/Infrared Array Camera (IRAC) photometry and model atmospheres to account for the emergent flux at other wavelengths.  Interestingly, they measured a value of $\log (L_\mathrm{bol}/L_\odot) = -6.13^{+0.20}_{-0.16}$ which is significantly brighter than our value, even when taking into account the uncertainties in our respective measurements.  Our bolometric flux measurements agree within the uncertainties and so the difference in bolometric luminosities can be almost  completely explained by the difference in parallaxes used to compute the distance to \WISEeighteentwentyeight, 70$\pm$14 mas in the case of  \citeauthor{2013Sci...341.1492D} and 100.7$\pm$2.3 mas in the case of \citeauthor{2019ApJS..240...19K}.  The relative uncertainty in  \citeauthor{2019ApJS..240...19K}  parallax is a factor of 9 lower than the \citeauthor{2013Sci...341.1492D} parallax so hereafter we will use our value of $\log (L_\mathrm{bol}/\mathcal{L}^N_\odot)$.

\subsection{Effective Temperature}

Estimating the effective temperature of \WISEeighteentwentyeight\ has proven difficult given its intrinsic faintness and contaminated spectrum.  \citet{2011ApJ...743...50C} estimated the effective temperature to be less than 300 K based on the near equal height of the spectral peaks in the $J$ and $H$ bands and its extreme $J-W2$ color while \citet{2013ApJ...763..130L} suggest that \WISEeighteentwentyeight\ is actually a $\sim$2 Gyr old binary with $M$=10, 7 $M_\mathrm{Jup}$, \teff=325, 300 K and \logg=4.5, 4.0 [ cm s$^{-2}$]. \citet{2013ApJ...764..101B} fit multi-band absolute photometry in various combinations and found \teff $\sim$ 250--450 K but noted that the ``estimated physical parameters should be taken with a grain of salt'' given how poor the models fit the data.  \citet{2013Sci...341.1492D} used their bolometric luminosity measurement (see \S\ref{sec:lbol}) and evolutionary models to estimate an effective temperature of 470 to 640 K for ages between 1 and 5 Gyr, but this rather high value is a result of their relatively low-precision parallax measurement. 

We can make a zeroth-order estimate of the effective temperature of \WISEeighteentwentyeight\ by making the reasonable assumption that its radius is $R\approx 1 \mathcal{R}_\textrm{J}^\textrm{N}$ -  the nominal value for Jupiter's equitorial radius of 7.1492$\times 10^7$ m \citep{2015arXiv151007674M} -  since the radii of all evolved brown dwarfs are with $\sim$30\% of this value due to the competing effects of Coulomb and electron degeneracy effects \citep{2001RvMP...73..719B}. With \teff $= (d^2f_\textrm{bol}/\sigma R^2)^{-1/4}$, where $\sigma$ is the Stefan-Boltzmann constant, we find \teff=426 K.  If \WISEeighteentwentyeight\ is an equal-luminosity binary, the effective temperature of each component would be \teff=351 K.  Both values are consistent with previous estimates (with the exception of the \citeauthor{2013Sci...341.1492D} estimate which is result of their large bolometric luminosity measurement) confirming that \WISEeighteentwentyeight\ is still one of the coolest brown dwarfs known, especially if it is a binary.  In the following two sections, we make additional estimates of the effective temperature of \WISEeighteentwentyeight\ using both evolutionary and atmospheric models.

\subsubsection{Bolometric Luminosity \& Evolutionary Models}

\begin{figure}
 \centerline{\hbox{\includegraphics[width=3.5in,angle=0]{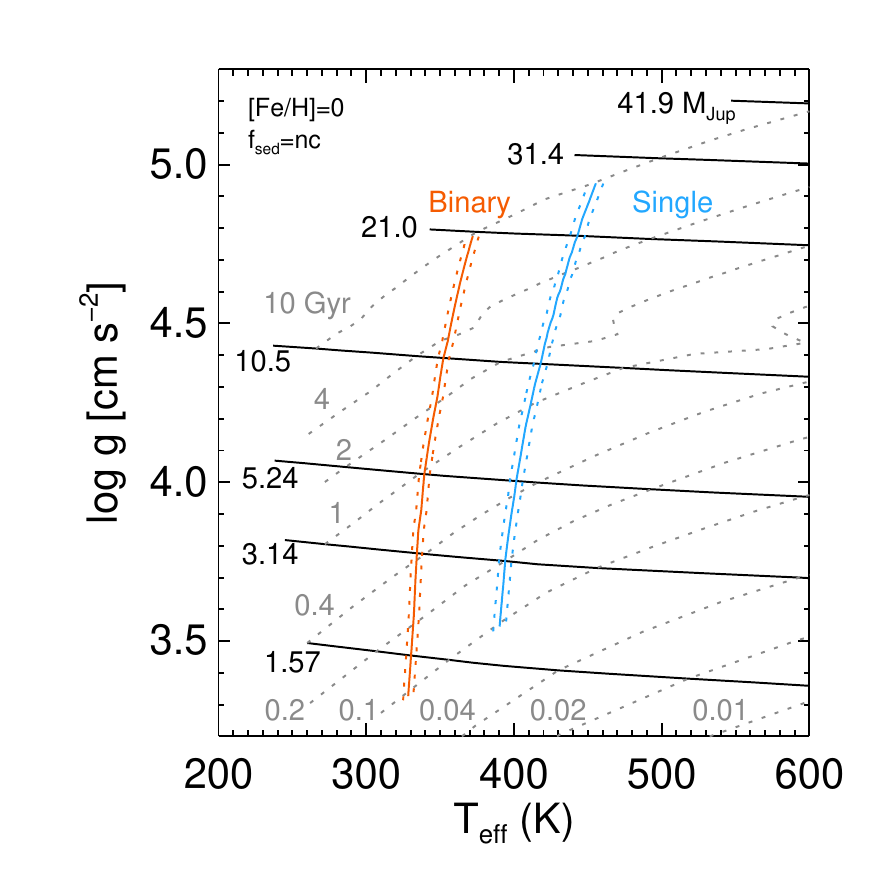}}} \caption{\label{fig:evolbol} Evolution of Bobcat Sonora solar metallicity cloudless brown dwarfs in the effective temperature surface gravity plane \citep[][Marley et al., submitted]{marley_mark_2020_3733843}.  The black lines are cooling tracks for brown dwarfs with masses of 41.9, 31.4, 21, 10.5, 5.24, 3.14, and 1.57 $\mathcal{M}^N_\textrm{J}$ while the grey lines are isochrones for ages of 10, 4, 2, 1, 0.4, 0.2, 0.1, 0.04, 0.02, and 0.01 Gyr.  The locus of points with bolometric luminosities equal to that of \WISEeighteentwentyeight\ for ages between 0.1 and 10 Gyr are shown as the solid near-vertical lines while the $\pm 1\sigma$ uncertainties on the bolometric luminosities are shown as dotted lines.}
\end{figure}

With an estimate of \WISEeighteentwentyeight's bolometric luminosity, we can use evolutionary models to estimate its effective temperature following the method described by \citet{2000ApJ...541..374S}.  Figure \ref{fig:evolbol}  shows the evolution of solar metallicity cloudless brown dwarfs in the effective temperature/surface gravity plane (Marley et al., 2020).  The locus of points with bolometric luminosities equal to that of \WISEeighteentwentyeight\ for ages between 0.1 and 10 Gyr is shown as a near-vertical blue line which constrains its effective temperature to lie between 386 $\leq$ \teff $\leq$ 461 K.  This is significantly lower than the value found by \citet{2013Sci...341.1492D} using a similar technique, but as noted previously this is simply a result of using a different parallax value to compute the bolometric luminosity.  

The near-vertical red line corresponds to the locus of points with bolometric luminosities equal to half that of \WISEeighteentwentyeight\ which is appropriate if it is an equal-brightness binary.  In this case, the effective temperature of the two components must lie between 325 $\leq$ \teff $\leq$ 376 K which is more consistent with previous estimates based on analyses of the colors and spectral energy distribution of \WISEeighteentwentyeight.

\subsubsection{Atmospheric Models}
\label{sec:atmosfits}

We also estimate the effective temperature of \WISEeighteentwentyeight\ by fitting the \textsl{HST} spectrum and spectral energy distribution shown in Figure \ref{fig:SED} with two grids of atmospheric models:  the \citet{2014ApJ...787...78M} models that include the formation of water clouds and the cloudless \citet{marley_mark_2018_1309035} Sonora Bobcat models.  The solar metallicity, partly cloudy (\fsed=5, $h$=0.5) \footnote{\fsed\ is the sedimentation parameter which describes the efficiency of condensate sedimentation \citep{2001ApJ...556..872A} and $h$ is a parameter that ranges from 0 to 1 that describes the fraction of the atmosphere that can be described by a cloud-free column \cite{2010ApJ...723L.117M}.  Larger values of $f_\textrm{sed}$ imply thinner clouds and a value of $h=0.5$ implies a 50\% cloud coverage.} \citeauthor{2014ApJ...787...78M} grid is a function of two free parameters, $\thetaatm=\{\teff, g\}$, with \teff=200--450 K in steps of 25 K (except for \teff=425 K), $\log g$=3.5, 4.0, 4.5, 5.0 [cm s$^{-2}$], for a total of 40 models.  The cloudless (\fsed=nc) Sonora Bobcat grid is a function of four free parameters, $\thetaatm=\{\teff, g, [\textrm{M/H}], \textrm{[C/O]}\}$, with \teff=250--450 K in steps of 50 K, \logg = 4.0, 4.5 [ cm s$^{-2}$], [M/H] = $+$0.5, 0.0, $-$0.5, and [C/O] = 0.0, $-$0.3, $-$0.6 \footnote{The model file naming convention is that the solar C/O ratio is denoted as C/O=1 which we write here as [C/O]=0.0.} for a total of 72 models.  We note that the [M/H] and [C/O] values are for the bulk composition of the atmosphere and do not necessarily reflect the composition of the gas phase because condensate species such as enstatite (MgSiO$_3$) and forsterite (Mg$_2$SiO$_4$) form at these temperatures and pressures.

One-dimensional model atmospheres provide flux densities at the surface of the brown dwarf at $n$ discrete wavelengths, i.e. $\mathbfcal{F}_\lambda(\thetaatm)=[\mathcal{F}_\lambda(\thetaatm, \lambda_1), \mathcal{F}_\lambda(\thetaatm, \lambda_2), \mathcal{F}_\lambda(\thetaatm, \lambda_3), ..., \mathcal{F}_\lambda(\thetaatm, \lambda_n)]$.  In order to compare each of the $j$ models in a given grid to the observations of \WISEeighteentwentyeight, we must simulate observing the model spectra with the techniques used to collect our data.  The spectroscopic portion of the data is given by,

\begin{equation}
\mathbfcal{M}_{\lambda}^{HST}(\thetaatmx{j}) = h(\lambda) \ast K(\lambda) \ast \mathbfcal{F}_\lambda(\thetaatmx{j}),   
\end{equation}
\noindent
where $K(\lambda$) is a Gaussian kernel that smooths the $j$th model $\mathbfcal{F}_\lambda(\thetaatmx{j})$ to the resolving power of the data, $h(\lambda)$ is a resampling kernel that resamples the smoothed model spectrum onto the wavelength grid of the data, and $'*'$ denotes a convolution.  The resolving powers of the \textit{HST} spectrum are given in \S\ref{sec:obs} and we used a linear interpolation kernel for $h(\lambda$).  

The photometric portions of the data can be simulated by integrating the model spectra over the appropriate bandpasses.  The MKO $K$-band point is given by,

\begin{equation}
\label{eq:MKO}
 \mathcal{M}^{K}_\lambda(\thetaatmx{j}) = \frac{\int \lambda \mathcal{F}_\lambda (\thetaatmx{j}, \lambda) S(\lambda)\, d\lambda}{\int \lambda S(\lambda)\, d\lambda},
\end{equation}
\noindent
where $S(\lambda)$ is the system response function of the $K$-band which we assume to be given by the product of the filter transmission and the typical atmospheric transmission at an air mass of 1.  The $\lambda$ inside the integral converts the energy flux densities to photon flux densities which ensures that the integrated fluxes are proportional to the observed photon count rate.  The [3.6] and [4.5] IRAC points are given by,

\begin{equation}
 \mathcal{M}^{\mathrm{IRAC}}_{\nu}(\thetaatmx{j}) = \frac{\int (\nu_{0} / \nu) \mathcal{F}_{\nu}(\thetaatmx{j},\nu) S(\nu)\, d\nu}{\int ( \nu_{0} / \nu)^{2}S(\nu)\, d\nu},
\end{equation}
\noindent
where $\nu_0$ is the nominal frequency, and where $S(\nu)$ is the system response function of the telescope plus instrument plus detector system in units of e$^-$ per photon \citep{2006ApJ...648..614C}.   The per-frequency flux density is then converted to a per-wavelength flux density $\mathcal{M}^{\mathrm{IRAC}}_{\lambda}(\thetaatmx{j})$ at the nominal wavelengths of the bandpasses.  Finally, the \WISE\ $W1$ and $W2$ points are given by,

\begin{equation}
 \mathcal{M}^{\mathrm{WISE}}_{\lambda}(\thetaatmx{j}) = \frac{\int \mathcal{F}_\lambda (\thetaatmx{j}, \lambda) S(\lambda)\, d\lambda}{\int S(\lambda)\, d\lambda}
\end{equation}
\noindent
where $S(\lambda)$ is the system response functions.  Note that we do not include the factor of $\lambda$ found in the integral like in Equation \ref{eq:MKO} because the filter transmission curves have already been multiplied by $\lambda$.  The $j$th model at the surface of the brown dwarfs is then given by,

\begin{align}
  \mathbfcal{M}(\thetaatmx{j}) =  & [\mathbfcal{M}_\lambda^{HST}(\thetaatmx{j}),\mathcal{M}^{K}_\lambda(\thetaatmx{j}), \\
& \mathcal{M}^{[3.6]}_\lambda(\thetaatmx{j}), \mathcal{M}^{[4.5]}_\lambda(\thetaatmx{j}), \\
& \mathcal{M}^{W2}_\lambda(\thetaatmx{j})]. 
\end{align}

\begin{figure*}[h!]
 \centerline{\hbox{\includegraphics[width=6in,angle=0]{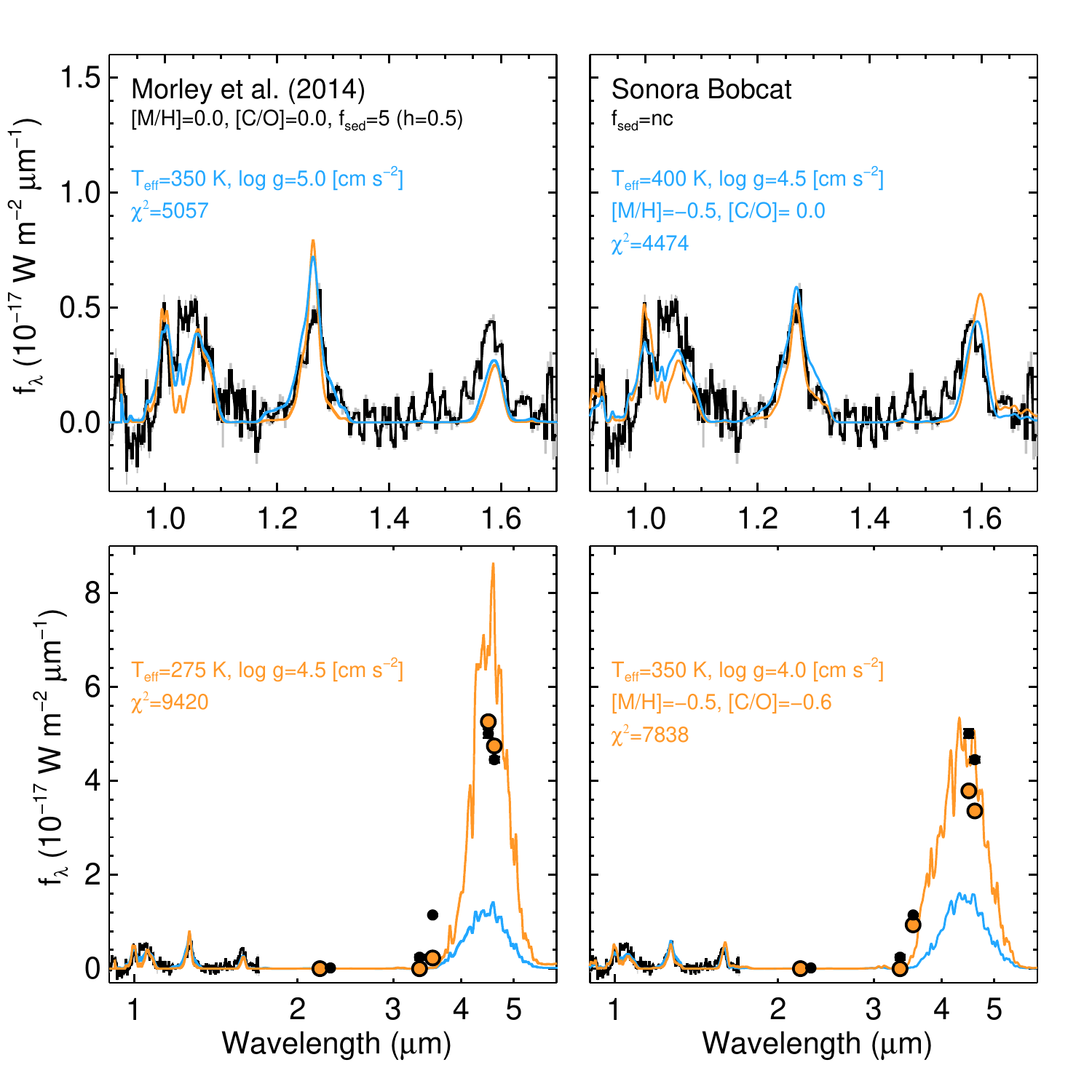}}} \caption{\label{fig:specfits} Best fitting \cite{2014ApJ...787...78M} model spectra (\textit{left column}) and Sonora models (\textit{right column}).  The top row shows the fits to the \textit{HST} spectrum while the bottom row shows the fits to the spectral energy distribution.  The flux densities of \WISEeighteentwentyeight\ in the $K$-band, $W1$, [3.6], [4.5], and $W2$ are shown as filled black circles while the flux densities of the models integrated over the corresponding bandpasses are shown as filled orange circles.  The $K$-band model flux density has been slightly offset in wavelength for clarity.  For comparison, the best fitting model spectra to the \textit{HST} spectrum are shown in the bottom panels and the best fitting model spectra to the spectral energy distribution are shown in the top panels.}
\end{figure*}

In order to compare a model spectral energy distribution $\mathbfcal{M}(\thetaatmx{j})$ to the data, we first need to multiple it by $(R/d)^2$, where $R$ is the radius of the brown dwarf and $d$ is its distance, in order to convert the flux densities at the surface of the brown dwarf to flux densities observed on or near Earth.  While the distance to \WISEeighteentwentyeight\ is known to be 9.93 pc \citep{2019ApJS..240...19K}, its radius is unknown and so we can only compare the relative shapes of the model spectra and data.  We identify the best fitting model spectrum in each grid following \citet{2008ApJ...678.1372C} where for each model $j$, we compute a $\chi^2$ statistic,

\begin{equation}
\chi^2_j = \sum_{i=1}^n \left [ \frac{f_i - C_j\mathcal{M}_i(\thetaatmx{j})}{s_i} \right ]^2,
\end{equation}

\noindent
where $n$ is the number of data points, $f_i$ and $\mathcal{M}_{j,i}$ are the flux densities of the data and $j$th model respectively, $s_i$ is the uncertainty in the $i$th data point\footnote{The standard deviation in the $\chi^2$ statistic requires knowledge of the uncertainty in our model which is beyond the scope of this work.  We therefore assume that the model standard deviations are given by the uncertainties on our data points.}
, and $C_j$ is an unknown multiplicative constant.  For each model $j$, we determine the value of $C_j$ by minimizing $\chi^2$ with respect to $C_j$ to give,

\begin{equation}
C_j = \frac{\sum_i f_i \mathcal{M}_{j,i}/s_i^2}{\sum_i \mathcal{M}_{j,i}/s_i^2}.
\end{equation}
\noindent
The best fitting model in each grid is that with the lowest $\chi^2$ value.  We fit both grids of model spectra to the \textit{HST} spectrum and then to the entire spectral energy distribution (the \textit{HST} spectrum and photometry) and the results are given in columns 1--8 in Table \ref{tab:parms} and shown in Figure \ref{fig:specfits}.

\pagebreak

\movetabledown=1.5in 
\begin{rotatetable}
\begin{deluxetable*}{llcccccccccccccc}
\label{tab:parms}
\tablecolumns{1}
\tabletypesize{\scriptsize} 
\tablecaption{\label{tab:properties}WISE 1828$+$2650 Model Fits}
\tablehead{
\colhead{} & 
\multicolumn{6}{c}{Best Fit Atmospheric Parameters \tablenotemark{a}} &
\colhead{} & 
\multicolumn{4}{c}{Evolution Parameters \tablenotemark{b}} &
\colhead{} & 
\multicolumn{2}{c}{Estimated\tablenotemark{c}} \\
\cline{2-7}
\cline{9-12}
\cline{14-16}
\colhead{Models} &
\colhead{\teff} & 
\colhead{$\log g$} & 
\colhead{[Fe/H]} & 
\colhead{[C/O]} &
\colhead{$C$} & 
\colhead{$\chi^2$/dof} & 
\colhead{} &
\colhead{$M$} &
\colhead{$R$} &
\colhead{$\tau$} &
\colhead{$\log L/L_\odot$} &
\colhead{} &
\colhead{$d_C$} & 
\colhead{$R_C$} \\
\colhead{} &
\colhead{(K)} &  
\colhead{[cm s$^{-2}$]} & 
\colhead{} &
\colhead{} &
\colhead{} &
\colhead{} &
\colhead{} &
\colhead{$(\mathcal{M}_\textrm{J}^\textrm{N})$} &
\colhead{$(\mathcal{R}_\textrm{J}^\textrm{N})$} &
\colhead{(Gyr)} &
\colhead{} &
\colhead{} &
\colhead{(pc)} & 
\colhead{$(\mathcal{R}_\textrm{J}^\textrm{N})$} \\
\colhead{(1)} &
\colhead{(2)} &
\colhead{(3)} &
\colhead{(5)} &
\colhead{(6)} &
\colhead{(7)} &
\colhead{(8)} &
\colhead{} &
\colhead{(9)} &
\colhead{(10)} &
\colhead{(11)} &
\colhead{(12)} &
\colhead{} &
\colhead{(13)}  & 
\colhead{(14)} 
}

\startdata 
\multicolumn{15}{c}{\textit{HST} Spectrum} \\
\hline 
Morley et al. (2014) & 350     & 5.0     &  0.0\tablenotemark{d}     & 0.0\tablenotemark{d}     & 2.7070$\times 10^{-20}$ & 5058/211 & & 29.6      & 0.85      & 26.7\tablenotemark{e}      & $-$6.98 & & 12.0 & 0.71 \\
Sonora Bobcat              & 400     & 4.5     & $-$0.5                    & 0.0                      & 1.6848$\times 10^{-20}$ & 4474/211 & & 13.0      & 1.00      & 3.14      & $-$6.61 & & 17.9 & 0.56 \\
\hline 
\multicolumn{15}{c}{Spectral Energy Distribution} \\
\hline 
Morley et al. (2014) & 275     & 4.5     &  0.0\tablenotemark{d}     & 0.0\tablenotemark{d}     &4.8680$\times 10^{-19}$ & 9421/216 & & 12.3      & 0.98      & 12.3      & $-$7.29 & & 3.24  & 2.99 \\
Sonora Bobcat              & 350     & 4.0     & $-$0.5                    & $-$0.6                   & 9.8445$\times 10^{-20}$ & 7839/216 & & 4.98      & 1.10      & 0.760      & $-$6.76 & & 8.16  & 1.34 \\
\hline 
\multicolumn{15}{c}{\textit{HST} Spectrum (Binary Fit)\tablenotemark{e}} \\
\hline 
Morley et al. (2014) & 350,350 & 5.0,5.0 &  0.0,0.0\tablenotemark{d} & 0.0,0.0\tablenotemark{d} & 3.6496$\times 10^{-36}$ & 5058/211 & & 29.6,29.6 & 0.85,0.85 & 26.7,26.7\tablenotemark{e} & $-$6.98,$-$6.98 & & 17.0  & $\cdots$ \\
Sonora Bobcat              & 400,400 & 4.5,4.5 & $-$0.5,$-$0.5             & 0.0,0.0                  & 1.6360$\times 10^{-36}$ & 4474/211 & & 13.0,13.0 & 1.00,1.00 & 3.14,3.14 & $-$6.61,$-$6.61 & & 25.3  & $\cdots$ \\
\hline 
\multicolumn{15}{c}{Spectral Energy Distribution (Binary Fit)\tablenotemark{e}} \\
\hline 
Morley et al. (2014) & 275,275 & 4.5,4.5 &  0.0,0.0\tablenotemark{d} & 0.0,0.0\tablenotemark{d} & 4.9947$\times 10^{-35}$  & 9421/216 & & 12.3,12.3 & 0.98,0.98 & 12.3,12.3 & $-$7.29,$-$7.29 & & 4.59  & $\cdots$ \\
Sonora Bobcat              & 300,350 & 4.0,4.0 & 0.0,$-$0.5             & $-$0.3,$-$0.6            & 1.2819$\times 10^{-35}$ & 7286/216 & & 4.65,4.75 & 1.09,1.10 & 1.41,0.760 & $-$7.04,$-$6.76 & & 9.03 & $\cdots$ \\
\enddata 
\tablenotetext{a}{The Morley et al. models have \fsed=5, h=0.5 while the Sonora Bobcat models have \fsed=nc.} 
\tablenotetext{b}{The values for $M$, $R$, and $\tau$ are computed using the estimated atmospheric parameters \thetaatm\ and the Sonora Bobcat evolutionary models. The models computed with the same metallicity as the derived atmospheric parameters were used.   $\mathcal{M}_\textrm{J}^\textrm{N}$ is the nominal Jupiter mass (assuming $G=6.67430\times 10^{-11}$ m$^3$ kg$^{-1}$ s$^{-2}$) while $\mathcal{R}_\textrm{J}^\textrm{N}$ is the nominal value for Jupiter's equitorial radius of 7.1492$\times 10^7$ m \citep{2015arXiv151007674M}.}
\tablenotetext{c}{The values for $d_C$ is estimated as $d_C=R(\thetaatm)/\sqrt{C}$ while the value for $R_C$ is estimated as $R_C=\sqrt{C}d$ where $d$ is the observed distance to WISE 1828$+$26 of 9.93 pc \citep{2019ApJS..240...19K} and $\mathcal{R}^\textrm{N}_\odot=6.957 \times 10^8$ m. \citep{2015arXiv151007674M}.} 
\tablenotetext{d}{Not a free parameter in the fit.} 
\tablenotetext{e}{This value is an extrapolation beyond the 20 Gyr limit of the evolution calculations.} 

\end{deluxetable*}
\end{rotatetable}

\clearpage

\begin{figure*}[t!]
 \centerline{\hbox{\includegraphics[width=6in,angle=0]{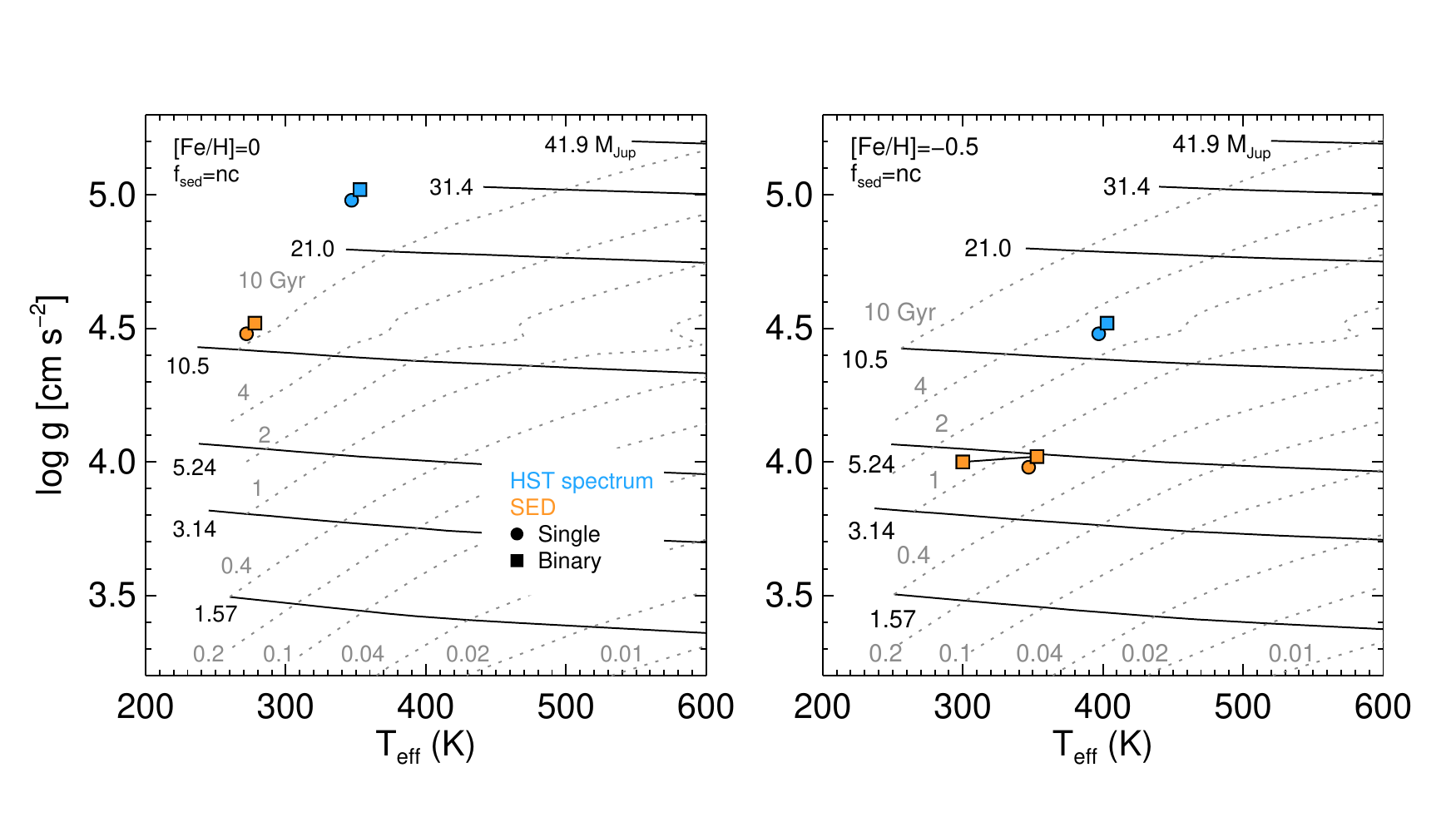}}} \caption{\label{fig:evoatmo}Evolution of solar metallicity (\textit{left}) and [Fe/H]=$-$0.5 (\textit{right}) cloudless brown dwarfs in the effective temperature surface gravity plane \citep[][Marley et al., submitted]{marley_mark_2020_3733843}.  The black lines are cooling tracks for brown dwarfs with masses of 41.9, 31.4, 21, 10.5, 5.24, 3.14, and 1.57 $\mathcal{M}^N_\textrm{J}$ while the grey lines are isochrones for ages of 10, 4, 2, 1, 0.4, 0.2, 0.1, 0.04, 0.02, and 0.01 Gyr.  Also plotted are the results of the atmospheric model fits discussed in \S\ref{sec:atmosfits}.  The results of fits assuming \WISEeighteentwentyeight\ is a single object are shown as circles while the results assuming it is binary are shown as squares.  The results of fits to just the \textit{HST} spectrum are shown in blue while the results for the spectral energy distribution are shown in orange. Finally, the points have been shifted slightly from their true values or visual clarity.}
\end{figure*}

The atmospheric parameters of the best-fit model spectrum can be used to infer additional properties of \WISEeighteentwentyeight\ when they are combined with evolutionary models because the models provide a unique mapping between the structural parameters of radius ($R$), mass ($M$), age ($\tau$), and luminosity ($L_\textrm{bol}$)  and the atmospheric parameters of effective temperature and surface gravity and so columns 9--11 in Table \ref{tab:parms} gives the mass, radius, age, and luminosity of \WISEeighteentwentyeight\ according to the Bobcat Sonora evolutionary models \citep[][Marley et al., submitted]{marley_mark_2020_3733843}.  The scale factor $C$ is equal to $(R/d)^2$ and so with the evolutionary model radius $R(\thetaatm)$ we can empirically estimate the distance to \WISEeighteentwentyeight\ and compare it to the known distance of $d$=9.93$\pm$0.23 pc \citep{2019ApJS..240...19K}.  Alternatively, we can use the known distance to estimate its radius which can then be compared to the predictions of evolutionary models.  These estimates are given in columns 13 and 14 of Table \ref{tab:parms}.  The location of the best-fit values in the effective temperature/surface gravity plane with respect to the evolution of cloudless brown dwarfs is also shown in Figure \ref{fig:evoatmo}.

\begin{figure*}
 \centerline{\hbox{\includegraphics[width=6in,angle=0]{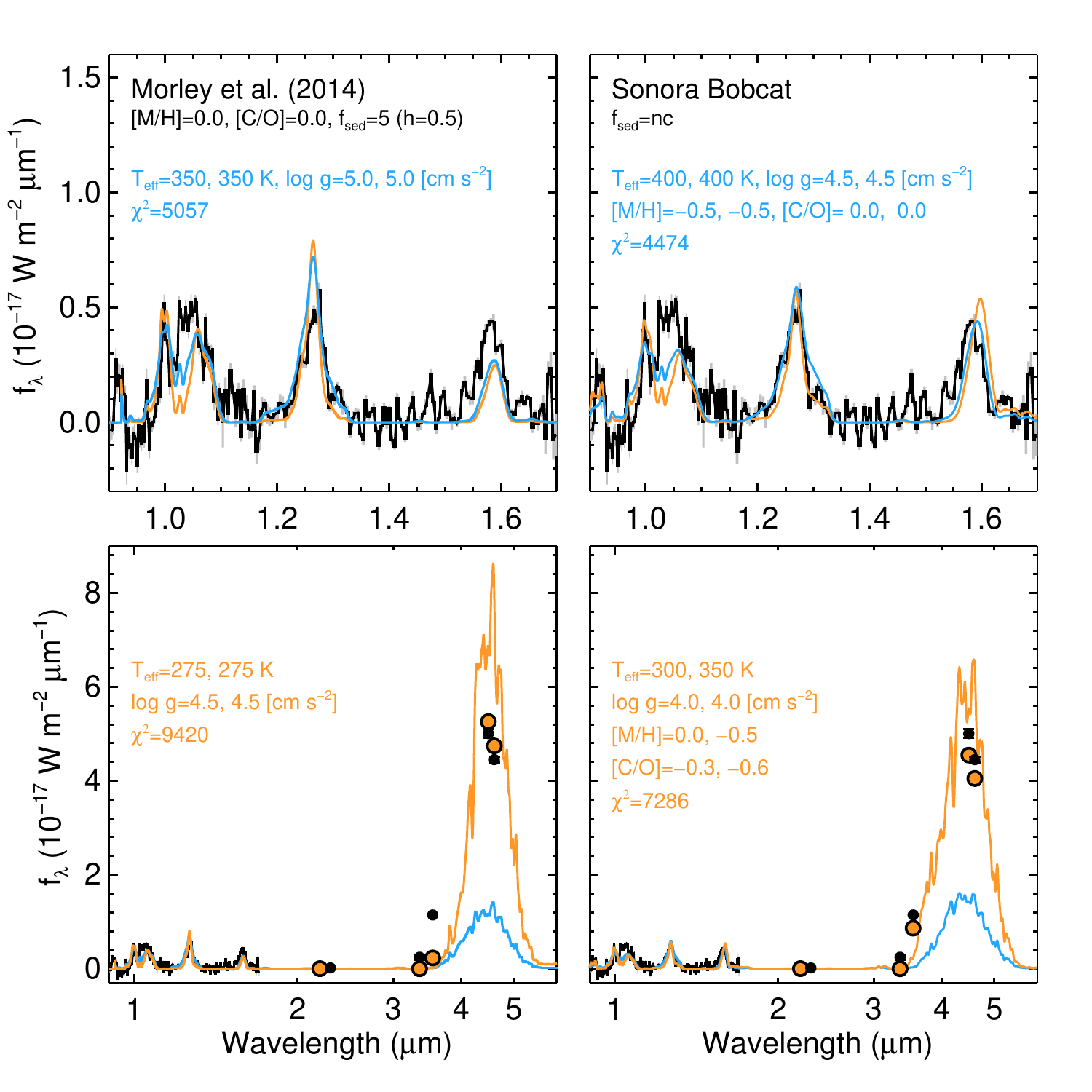}}} \caption{\label{fig:binaryfit} Best fitting \cite{2014ApJ...787...78M} binary model spectra (\textit{left column}) and Sonora Bobcat binary model spectra (\textit{right column}).  Otherwise similar to Figure \ref{fig:specfits}.}
\end{figure*}

Given that \WISEeighteentwentyeight\ is overluminous in nearly all color magnitude diagrams,  we also fit the data to model binary stars.  For each grid, we simulated a collection of binary systems by scaling each of the model spectra by $R(\thetaatm)^2$ (the radius given by the evolutionary models that corresponds to $\thetaatm$) and then summed pairs of models ($k,l$) as:

\begin{align}
  \mathbfcal{M}_j = &  R(\thetaatmx{k})^2[\mathbfcal{M}_{\lambda}^{HST}(\thetaatmx{k}),\mathcal{M}^{K}_{\lambda}(\thetaatmx{k}), \\
& \mathcal {M}^{[3.6]}_{\lambda}(\thetaatmx{k}), \mathcal{M}^{[4.5]}_{\lambda}(\thetaatmx{k}), \\
& \mathcal{M}^{W2}_{\lambda}(\thetaatmx{k})]\,\,\, + \\
 & R(\thetaatmx{l})^2[\mathbfcal{M}_{\lambda}^{HST}(\thetaatmx{l}),\mathcal{M}^{K}_{\lambda}(\thetaatmx{l}),\\
& \mathcal {M}^{[3.6]}_{\lambda}(\thetaatmx{l}), \mathcal{M}^{[4.5]}_{\lambda}(\thetaatmx{l}),\mathcal{M}^{W2}_{\lambda}(\thetaatmx{l})].
\end{align}
However, we placed the following constraints, 
\begin{align}
 |\tau_k - \tau_l| \leq & 1 \textrm{ Gyr}, \\
 |\textrm{[Fe/H]}_k - \textrm{[Fe/H]}_l| \leq & 0.5 \textrm{ dex}, \\
 |\textrm{[C/O]}_k - \textrm{[C/O]}_l| \leq & 0.3 \textrm{ dex},
\end{align}

\noindent
on the components of the binary to ensure that they have similar compositions and ages.  While it is reasonable to assume that the components of a binary have identical ages and compositions, the course sampling of the atmospheric parameters requires weakening this constraint in order to avoid creating only equal-parameter binaries.  The results of these fits are given in columns 1--8 in Table \ref{tab:parms} and shown in Figures \ref{fig:evoatmo} and \ref{fig:binaryfit}.  The corresponding structural parameters are given in columns 9--12. Since the model spectra are first scaled by $R(\thetaatm)^2$, the scale factor is simply given by $C=(1/d)^2$ and so we can provide an empirical estimate of the distance as $d_C=1/\sqrt{C}$, and the results are given in column 13 of Table \ref{tab:parms}. 

\section{Discussion}

Statistically speaking, all of the model spectra are poor fits to the data since the $\chi^2$ values of the fits are factors of 20--40 times larger than the number of degrees freedom.  Given the complexity of the models atmospheres, statistically poor fits to brown dwarf spectra are not uncommon, especially when the observations span a broad wavelength range \citep[e.g.,][]{2008ApJ...678.1372C,2009ApJ...702..154S,2015ApJ...804...92S,2016AJ....152...78L,2018ApJ...858...97M}.  As a result, we do not attempt to provide statistical uncertainties on the derived parameters.   Nevertheless, it is still instructive to take the fits at face value because they not only provide some insight into the physical properties of the brown dwarfs but they also allow us to identify wavelengths where the models need improvement.

The \textit{HST} spectrum is best fit by the Sonora model spectrum (upper right panel of Figure \ref{fig:specfits}) because it more accurately reproduces the heights of the $J$- and $H$- band peaks.  Both models fit the $Y$-band peak poorly because the NH$_3$ band in both models is too strong.  This is most likely a result of the fact that the atmospheric models assume equilibrium chemistry and it has been shown that vertical mixing within an atmosphere keeps the nitrogen chemistry from coming into equilibrium such that the abundance of N$_2$ is increased relative to equilibrium and the abundance of NH$_3$ is decreased relative to equilibrium \citep[e.g.,][]{2003IAUS..211..345S,2006ApJ...647..552S}.  The presence of this NH$_3$ band also makes it difficult to assess whether they exhibit the 1.015 $\mu$m CH$_4$ band that we tentatively identified in the data in \S\ref{sec:thespectrum}.  

The derived \teff\, and \logg\ values, 350 K/5.0 [cm s$^{-2}$] and 400 K/4.5 [cm s$^{-2}$], for the \citeauthor{2014ApJ...787...78M} (solar composition, partly cloudy) and Sonora Bobcat models (cloud-free, with varied metallicity and C/O) respectively, are similar.  The corresponding age of 26.7 Gyr for the \citeauthor{2014ApJ...787...78M} model (a result of the high surface gravity of $\log g = 5$) is clearly inconsistent with the age of the Universe.  This underscores the fact that care must be taken when deriving the atmospheric properties of Y dwarfs because while atmospheric models can be constructed with various effective temperatures and surface gravities, objects with these atmospheres may not actually exist in our Universe \citep[see also][]{2015ApJ...804...92S}. The Sonora Bobcat model gives a more realistic age of 3.14 Gyr, but either has a large disparity between the $d_C$ values of 17.9 pc and the known distance of 9.93$\pm$0.23 pc or an unphysical $R_C$ of 0.56 $\mathcal{R}_\mathrm{J}^\mathrm{N}$. 

Turning to the fits of the full spectral energy distribution shown in the bottom row of Figure \ref{fig:specfits}, we see that while both models generally reproduce the very red near- to mid-infrared color of \WISEeighteentwentyeight, the \cite{2014ApJ...787...78M} fail to reproduce the [3.6] point while the Sonora Bobcat models fail to reproduce the W2 and [4.5] points.  The \teff\, and \logg\, values of 275 K/4.5 [cm s$^{-2}$] for the \citeauthor{2014ApJ...787...78M} model fit result in model bolometric luminosity and $d_\mathrm{C}$ values that are completely inconsistent with observations and a large $R_\mathrm{C}$ that would require an unrealistically young age of a few Myr at these temperatures.  The Sonora Bobcat model matches the observed luminosity and distance better, which indicates that if \WISEeighteentwentyeight\ a single object, it has a mass of $\sim$5 $\mathcal{M}_\textrm{Jup}$, an age of 760 Myr, and both a subsolar metallicity and subsolar C/O ratio. 

\begin{figure}
 \centerline{\hbox{\includegraphics[width=3.5in,angle=0]{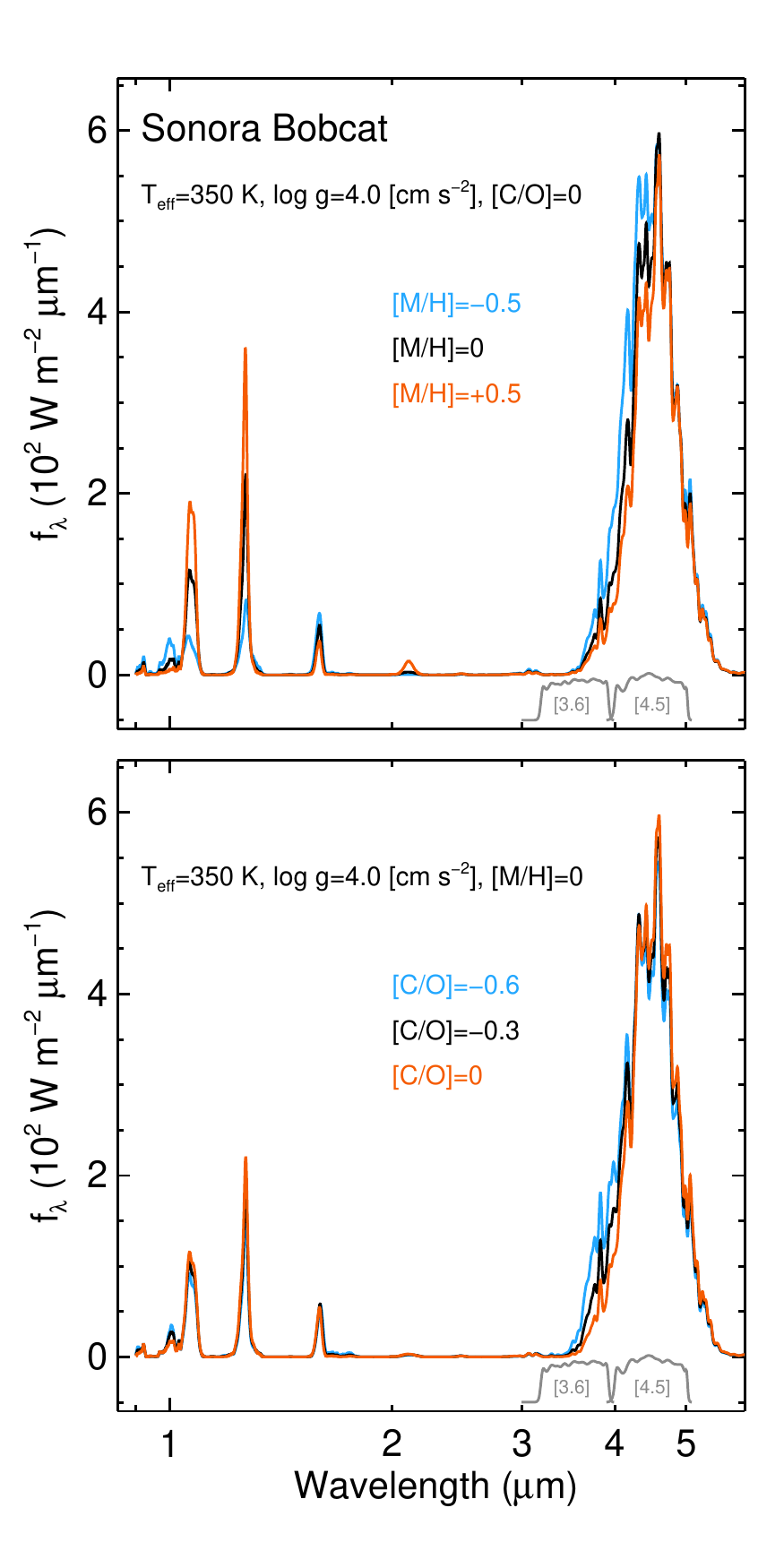}}} \caption{\label{fig:modseq} Sonora model spectra with atmospheric parameters typical for a Y dwarf (\teff=350 K, \logg=4.0 [cm s$^{-2}$] when [M/H] and [C/O] are varied. The models have been smoothed to $R$$\approx$180 and correspond to the emergent flux at the top of the atmosphere.  Also shown in grey are the transmission curves for the \textit{Spitzer}/IRAC [3.6] and [4.5] bands.}
\end{figure}

The mismatch at [3.6] between the data and the models that assume solar abundances is a well known problem \citep[e.g.,][]{2017ApJ...842..118L} and so the much better agreement between the [3.6] point and the Sonora Bobcat models is most likely a result of the fact that the Sonora models have [M/H] and [C/O] as free parameters, since the best fitting model has [M/H]=$-$0.5 and [C/O]=$-$0.6. Figure \ref{fig:modseq} show a sequence of model spectra with typical Y dwarf \teff\, and $g$ values but with variations in [M/H] and [C/O].  Changes in both metallicity and [C/O] affect the width of the flux peak at 4.5 $\mu$m in the sense that lower metallicity and lower [C/O] values result in a widening of the blue side of the emission peak.  While both the \textit{Spitzer} [3.6] and [4.5] bands sample this side of the peak, so little emission emerges from the model brown dwarf in the short-wavelength half of the [3.6] band that small changes in the emission peak results in a large change in the [3.6] magnitude.  Previous work by \citet{2018ApJ...858...97M} and \citet{2019ApJ...877...24Z} have also shown that subsolar [C/O] values are required in order to match the observations of Y dwarfs.  The aforementioned vertical mixing also impacts the carbon chemistry and thus the abundances of CO and CH$_4$.  Since the opacity of these molecules sculpt the 5 $\mu$m flux peak, non-equilibrium carbon chemistry can also have an impact on how well the \textit{Spitzer} and \textit{WISE} points are reproduced by the models \citep[e.g.,][]{2007ApJ...669.1248H,2020AJ....160...63M}.  Including both non-solar C/O ratios and non-equilibrium carbon chemistry in forward-model fits to Y dwarf observations may therefore be required in order to accurately reproduce the spectral morphology of the 5 $\mu$m flux peak.

Finally, we also plot the best fitting models to the \textit{HST} spectrum over the best fitting models to the spectral energy distribution and vice versa.  The differences between the fits at near-infrared wavelengths is minimal, but the \textsl{HST} spectrum fit fails to reproduce the dramatic rise in flux at $\sim5 \mu$m.  This  underscores the difficulty of estimating the atmospheric properties of Y dwarfs with datasets that have limited wavelength coverage \cite[see also][]{2015ApJ...804...92S}.

The fits to the binary models are shown in Figure \ref{fig:binaryfit}.  With the exception of the Sonora Bobcat fit to the entire spectral energy distribution (lower right panel), the binary models provide identical fits to those of the single models (the $d_C$ values of these models are a factor of $\sqrt{2}$ larger than the single fits and remain a poor match to the observed distance of 9.93 pc).  However, the Sonora Bobcat binary model fits the 5 $\mu$m peak much better than a single model and as a result, the $\chi^2$ falls from a value of 7838 to 7286.  Perhaps most importantly, $d_\mathrm{C}=9.03$ pc which is reasonably close to the measured value of 9.93.  As a result, a rough but consistent picture between the data and the atmospheric and evolutionary models emerges whereby \WISEeighteentwentyeight\ is a $\sim$1 Gyr old binary composed of two \teff $\sim$325 K, $\sim$5 $M_\textrm{Jup}$ brown dwarfs with subsolar [C/O] ratios.

\acknowledgments

This research is based on observations made with the NASA/ESA Hubble Space Telescope obtained from the Space Telescope Science Institute, which is operated by the Association of Universities for Research in Astronomy, Inc., under NASA contract NAS 5–26555. These observations are associated with program 12970.  This publication makes use of data products from the \textsl{Wide-field Infrared Survey Explorer}, which is a joint project of the University of California, Los Angeles, and the Jet Propulsion Laboratory/California Institute of Technology, and NEOWISE, which is a project of the Jet Propulsion Laboratory/California Institute of Technology. WISE and NEOWISE are funded by the National Aeronautics and Space Administration.  This research has benefitted from the Y Dwarf Compendium maintained by Michael Cushing at https://sites.google.com/view/ydwarfcompendium/.

\facilities{HST (WFC3), Spitzer (IRAC)}
\software{IDL}

\clearpage

\bibliographystyle{aasjournal}
\bibliography{/Users/michaelcushing/Science/Papers/ref,/Users/michaelcushing/Science/Papers/tmp}

\begin{thebibliography}{}
\expandafter\ifx\csname natexlab\endcsname\relax\def\natexlab#1{#1}\fi
\providecommand{\url}[1]{\href{#1}{#1}}
\providecommand{\dodoi}[1]{doi:~\href{http://doi.org/#1}{\nolinkurl{#1}}}
\providecommand{\doeprint}[1]{\href{http://ascl.net/#1}{\nolinkurl{http://ascl.net/#1}}}
\providecommand{\doarXiv}[1]{\href{https://arxiv.org/abs/#1}{\nolinkurl{https://arxiv.org/abs/#1}}}

\bibitem[{{Ackerman} \& {Marley}(2001)}]{2001ApJ...556..872A}
{Ackerman}, A.~S., \& {Marley}, M.~S. 2001, \apj, 556, 872

\bibitem[{{Bardalez Gagliuffi} {et~al.}(2020){Bardalez Gagliuffi}, {Faherty},
  {Schneider}, {Meisner}, {Caselden}, {Colin}, {Goodman}, {Kirkpatrick},
  {Kuchner}, {Gagn{\'e}}, {Logsdon}, {Burgasser}, {Allers}, {Debes},
  {Wisniewski}, {Rothermich}, {Andersen}, {Th{\'e}venot}, {Walla}, \& {Backyard
  Worlds: Planet 9 Collaboration}}]{2020ApJ...895..145B}
{Bardalez Gagliuffi}, D.~C., {Faherty}, J.~K., {Schneider}, A.~C., {et~al.}
  2020, \apj, 895, 145, \dodoi{10.3847/1538-4357/ab8d25}

\bibitem[{{Beichman} {et~al.}(2013){Beichman}, {Gelino}, {Kirkpatrick},
  {Barman}, {Marsh}, {Cushing}, \& {Wright}}]{2013ApJ...764..101B}
{Beichman}, C., {Gelino}, C.~R., {Kirkpatrick}, J.~D., {et~al.} 2013, \apj,
  764, 101, \dodoi{10.1088/0004-637X/764/1/101}

\bibitem[{{Bochanski} {et~al.}(2011){Bochanski}, {Burgasser}, {Simcoe}, \&
  {West}}]{2011AJ....142..169B}
{Bochanski}, J.~J., {Burgasser}, A.~J., {Simcoe}, R.~A., \& {West}, A.~A. 2011,
  \aj, 142, 169, \dodoi{10.1088/0004-6256/142/5/169}

\bibitem[{{Bowles} {et~al.}(2008){Bowles}, {Calcutt}, {Irwin}, \&
  {Temple}}]{2008Icar..196..612B}
{Bowles}, N., {Calcutt}, S., {Irwin}, P., \& {Temple}, J. 2008, \icarus, 196,
  612, \dodoi{10.1016/j.icarus.2007.12.029}

\bibitem[{{Burgasser} {et~al.}(2006){Burgasser}, {Geballe}, {Leggett},
  {Kirkpatrick}, \& {Golimowski}}]{2006ApJ...637.1067B}
{Burgasser}, A.~J., {Geballe}, T.~R., {Leggett}, S.~K., {Kirkpatrick}, J.~D.,
  \& {Golimowski}, D.~A. 2006, \apj, 637, 1067, \dodoi{10.1086/498563}

\bibitem[{{Burgasser} {et~al.}(2003){Burgasser}, {Kirkpatrick}, {Liebert}, \&
  {Burrows}}]{2003ApJ...594..510B}
{Burgasser}, A.~J., {Kirkpatrick}, J.~D., {Liebert}, J., \& {Burrows}, A. 2003,
  \apj, 594, 510, \dodoi{10.1086/376756}

\bibitem[{{Burningham} {et~al.}(2008){Burningham}, {Pinfield}, {Leggett},
  {Tamura}, {Lucas}, {Homeier}, {Day-Jones}, {Jones}, {Clarke}, {Ishii},
  {Kuzuhara}, {Lodieu}, {Zapatero Osorio}, {Venemans}, {Mortlock}, {Barrado Y
  Navascu{\'e}s}, {Mart{\'{\i}}n}, \& {Magazz{\`u}}}]{2008MNRAS.391..320B}
{Burningham}, B., {Pinfield}, D.~J., {Leggett}, S.~K., {et~al.} 2008, \mnras,
  391, 320, \dodoi{10.1111/j.1365-2966.2008.13885.x}

\bibitem[{{Burrows} {et~al.}(2001){Burrows}, {Hubbard}, {Lunine}, \&
  {Liebert}}]{2001RvMP...73..719B}
{Burrows}, A., {Hubbard}, W.~B., {Lunine}, J.~I., \& {Liebert}, J. 2001,
  Reviews of Modern Physics, 73, 719, \dodoi{10.1103/RevModPhys.73.719}

\bibitem[{{Burrows} {et~al.}(2003){Burrows}, {Sudarsky}, \&
  {Lunine}}]{2003ApJ...596..587B}
{Burrows}, A., {Sudarsky}, D., \& {Lunine}, J.~I. 2003, \apj, 596, 587

\bibitem[{{Cruikshank} \& {Binder}(1969)}]{1969CoLPL...6..275C}
{Cruikshank}, D.~P., \& {Binder}, A.~B. 1969, Communications of the Lunar and
  Planetary Laboratory, 6, 275

\bibitem[{{Cushing}(2014)}]{2014ASSL..401..113C}
{Cushing}, M.~C. 2014, in Astrophysics and Space Science Library, Vol. 401, 50
  Years of Brown Dwarfs, ed. V.~{Joergens}, 113

\bibitem[{{Cushing} {et~al.}(2006){Cushing}, {Roellig}, {Marley}, {Saumon},
  {Leggett}, {Kirkpatrick}, {Wilson}, {Sloan}, {Mainzer}, {Van Cleve}, \&
  {Houck}}]{2006ApJ...648..614C}
{Cushing}, M.~C., {Roellig}, T.~L., {Marley}, M.~S., {et~al.} 2006, \apj, 648,
  614, \dodoi{10.1086/505637}

\bibitem[{{Cushing} {et~al.}(2008){Cushing}, {Marley}, {Saumon}, {Kelly},
  {Vacca}, {Rayner}, {Freedman}, {Lodders}, \& {Roellig}}]{2008ApJ...678.1372C}
{Cushing}, M.~C., {Marley}, M.~S., {Saumon}, D., {et~al.} 2008, \apj, 678,
  1372, \dodoi{10.1086/526489}

\bibitem[{{Cushing} {et~al.}(2011){Cushing}, {Kirkpatrick}, {Gelino},
  {Griffith}, {Skrutskie}, {Mainzer}, {Marsh}, {Beichman}, {Burgasser},
  {Prato}, {Simcoe}, {Marley}, {Saumon}, {Freedman}, {Eisenhardt}, \&
  {Wright}}]{2011ApJ...743...50C}
{Cushing}, M.~C., {Kirkpatrick}, J.~D., {Gelino}, C.~R., {et~al.} 2011, \apj,
  743, 50, \dodoi{10.1088/0004-637X/743/1/50}

\bibitem[{{Cutri} {et~al.}(2013){Cutri}, {Wright}, {Conrow}, {Fowler},
  {Eisenhardt}, {Grillmair}, {Kirkpatrick}, {Masci}, {McCallon}, {Wheelock},
  {Fajardo-Acosta}, {Yan}, {Benford}, {Harbut}, {Jarrett}, {Lake}, {Leisawitz},
  {Ressler}, {Stanford}, {Tsai}, {Liu}, {Helou}, {Mainzer}, {Gettings},
  {Gonzalez}, {Hoffman}, {Marsh}, {Padgett}, {Skrutskie}, {Beck}, {Papin}, \&
  {Wittman}}]{2013wise.rept....1C}
{Cutri}, R.~M., {Wright}, E.~L., {Conrow}, T., {et~al.} 2013, {Explanatory
  Supplement to the AllWISE Data Release Products}, Tech. rep.

\bibitem[{{Dahn} {et~al.}(2002){Dahn}, {Harris}, {Vrba}, {Guetter}, {Canzian},
  {Henden}, {Levine}, {Luginbuhl}, {Monet}, {Monet}, {Pier}, {Stone}, {Walker},
  {Burgasser}, {Gizis}, {Kirkpatrick}, {Liebert}, \&
  {Reid}}]{2002AJ....124.1170D}
{Dahn}, C.~C., {Harris}, H.~C., {Vrba}, F.~J., {et~al.} 2002, \aj, 124, 1170,
  \dodoi{10.1086/341646}

\bibitem[{{Delorme} {et~al.}(2008{\natexlab{a}}){Delorme}, {Willott},
  {Forveille}, {Delfosse}, {Reyl{\'e}}, {Bertin}, {Albert}, {Artigau}, {Robin},
  {Allard}, {Doyon}, \& {Hill}}]{2008A&A...484..469D}
{Delorme}, P., {Willott}, C.~J., {Forveille}, T., {et~al.} 2008{\natexlab{a}},
  \aap, 484, 469, \dodoi{10.1051/0004-6361:20078843}

\bibitem[{{Delorme} {et~al.}(2008{\natexlab{b}}){Delorme}, {Delfosse},
  {Albert}, {Artigau}, {Forveille}, {Reyl{\'e}}, {Allard}, {Homeier}, {Robin},
  {Willott}, {Liu}, \& {Dupuy}}]{2008A&A...482..961D}
{Delorme}, P., {Delfosse}, X., {Albert}, L., {et~al.} 2008{\natexlab{b}}, \aap,
  482, 961, \dodoi{10.1051/0004-6361:20079317}

\bibitem[{{Delorme} {et~al.}(2010){Delorme}, {Albert}, {Forveille}, {Artigau},
  {Delfosse}, {Reyl{\'e}}, {Willott}, {Bertin}, {Wilkins}, {Allard}, \&
  {Arzoumanian}}]{2010A&A...518A..39D}
{Delorme}, P., {Albert}, L., {Forveille}, T., {et~al.} 2010, \aap, 518, A39,
  \dodoi{10.1051/0004-6361/201014277}

\bibitem[{{Dupuy} \& {Kraus}(2013)}]{2013Sci...341.1492D}
{Dupuy}, T.~J., \& {Kraus}, A.~L. 2013, Science, 341, 1492,
  \dodoi{10.1126/science.1241917}

\bibitem[{{Faherty} {et~al.}(2014){Faherty}, {Tinney}, {Skemer}, \&
  {Monson}}]{2014ApJ...793L..16F}
{Faherty}, J.~K., {Tinney}, C.~G., {Skemer}, A., \& {Monson}, A.~J. 2014,
  \apjl, 793, L16, \dodoi{10.1088/2041-8205/793/1/L16}

\bibitem[{{Filippazzo} {et~al.}(2015){Filippazzo}, {Rice}, {Faherty}, {Cruz},
  {Van Gordon}, \& {Looper}}]{2015ApJ...810..158F}
{Filippazzo}, J.~C., {Rice}, E.~L., {Faherty}, J., {et~al.} 2015, \apj, 810,
  158, \dodoi{10.1088/0004-637X/810/2/158}

\bibitem[{{Hubeny} \& {Burrows}(2007)}]{2007ApJ...669.1248H}
{Hubeny}, I., \& {Burrows}, A. 2007, \apj, 669, 1248, \dodoi{10.1086/522107}

\bibitem[{{Irwin} {et~al.}(1999){Irwin}, {Calcutt}, {Sihra}, {Taylor}, {Weir},
  {Ballard}, \& {Johnston}}]{1999JQSRT..62..193I}
{Irwin}, P.~G.~J., {Calcutt}, S.~B., {Sihra}, K., {et~al.} 1999, Journal of
  Quantitative Spectroscopy and Radiative Transfer, 62, 193,
  \dodoi{10.1016/S0022-4073(98)00103-4}

\bibitem[{{Irwin} {et~al.}(2010){Irwin}, {Teanby}, \&
  {Davis}}]{2010Icar..208..913I}
{Irwin}, P.~G.~J., {Teanby}, N.~A., \& {Davis}, G.~R. 2010, \icarus, 208, 913,
  \dodoi{10.1016/j.icarus.2010.03.017}

\bibitem[{{Jarrett} {et~al.}(2011){Jarrett}, {Cohen}, {Masci}, {Wright},
  {Stern}, {Benford}, {Blain}, {Carey}, {Cutri}, {Eisenhardt}, {Lonsdale},
  {Mainzer}, {Marsh}, {Padgett}, {Petty}, {Ressler}, {Skrutskie}, {Stanford},
  {Surace}, {Tsai}, {Wheelock}, \& {Yan}}]{2011ApJ...735..112J}
{Jarrett}, T.~H., {Cohen}, M., {Masci}, F., {et~al.} 2011, \apj, 735, 112,
  \dodoi{10.1088/0004-637X/735/2/112}

\bibitem[{{Kimble} {et~al.}(2008){Kimble}, {MacKenty}, {O'Connell}, \&
  {Townsend}}]{2008SPIE.7010E..43K}
{Kimble}, R.~A., {MacKenty}, J.~W., {O'Connell}, R.~W., \& {Townsend}, J.~A.
  2008, in Presented at the Society of Photo-Optical Instrumentation Engineers
  (SPIE) Conference, Vol. 7010, Society of Photo-Optical Instrumentation
  Engineers (SPIE) Conference Series

\bibitem[{{Kirkpatrick}(2008)}]{2008ASPC..384...85K}
{Kirkpatrick}, J.~D. 2008, in Astronomical Society of the Pacific Conference
  Series, Vol. 384, 14th Cambridge Workshop on Cool Stars, Stellar Systems, and
  the Sun, ed. {G.~van Belle}, 85

\bibitem[{{Kirkpatrick} {et~al.}(2011){Kirkpatrick}, {Cushing}, {Gelino},
  {Griffith}, {Skrutskie}, {Marsh}, {Wright}, {Mainzer}, {Eisenhardt},
  {McLean}, {Thompson}, {Bauer}, {Benford}, {Bridge}, {Lake}, {Petty},
  {Stanford}, {Tsai}, {Bailey}, {Beichman}, {Bloom}, {Bochanski}, {Burgasser},
  {Capak}, {Cruz}, {Hinz}, {Kartaltepe}, {Knox}, {Manohar}, {Masters},
  {Morales-Calder{\'o}n}, {Prato}, {Rodigas}, {Salvato}, {Schurr}, {Scoville},
  {Simcoe}, {Stapelfeldt}, {Stern}, {Stock}, \& {Vacca}}]{2011ApJS..197...19K}
{Kirkpatrick}, J.~D., {Cushing}, M.~C., {Gelino}, C.~R., {et~al.} 2011, \apjs,
  197, 19, \dodoi{10.1088/0067-0049/197/2/19}

\bibitem[{{Kirkpatrick} {et~al.}(2012){Kirkpatrick}, {Gelino}, {Cushing},
  {Mace}, {Griffith}, {Skrutskie}, {Marsh}, {Wright}, {Eisenhardt}, {McLean},
  {Mainzer}, {Burgasser}, {Tinney}, {Parker}, \&
  {Salter}}]{2012ApJ...753..156K}
{Kirkpatrick}, J.~D., {Gelino}, C.~R., {Cushing}, M.~C., {et~al.} 2012, \apj,
  753, 156, \dodoi{10.1088/0004-637X/753/2/156}

\bibitem[{{Kirkpatrick} {et~al.}(2019){Kirkpatrick}, {Martin}, {Smart},
  {Cayago}, {Beichman}, {Marocco}, {Gelino}, {Faherty}, {Cushing}, {Schneider},
  {Mace}, {Tinney}, {Wright}, {Lowrance}, {Ingalls}, {Vrba}, {Munn}, {Dahm}, \&
  {McLean}}]{2019ApJS..240...19K}
{Kirkpatrick}, J.~D., {Martin}, E.~C., {Smart}, R.~L., {et~al.} 2019, \apjs,
  240, 19, \dodoi{10.3847/1538-4365/aaf6af}

\bibitem[{{Kuntschner} {et~al.}(2011){Kuntschner}, {K{\"u}mmel}, {Walsh}, \&
  {Bushouse}}]{2011wfc..rept....5K}
{Kuntschner}, H., {K{\"u}mmel}, M., {Walsh}, J.~R., \& {Bushouse}, H. 2011,
  {Revised Flux Calibration of the WFC3 G102 and G141 grisms}, Tech. rep.

\bibitem[{{Lawrence} {et~al.}(2007){Lawrence}, {Warren}, {Almaini}, {Edge},
  {Hambly}, {Jameson}, {Lucas}, {Casali}, {Adamson}, {Dye}, {Emerson},
  {Foucaud}, {Hewett}, {Hirst}, {Hodgkin}, {Irwin}, {Lodieu}, {McMahon},
  {Simpson}, {Smail}, {Mortlock}, \& {Folger}}]{2007MNRAS.379.1599L}
{Lawrence}, A., {Warren}, S.~J., {Almaini}, O., {et~al.} 2007, \mnras, 379,
  1599, \dodoi{10.1111/j.1365-2966.2007.12040.x}

\bibitem[{{Leggett} {et~al.}(2007){Leggett}, {Marley}, {Freedman}, {Saumon},
  {Liu}, {Geballe}, {Golimowski}, \& {Stephens}}]{2007ApJ...667..537L}
{Leggett}, S.~K., {Marley}, M.~S., {Freedman}, R., {et~al.} 2007, \apj, 667,
  537, \dodoi{10.1086/519948}

\bibitem[{{Leggett} {et~al.}(2015){Leggett}, {Morley}, {Marley}, \&
  {Saumon}}]{2015ApJ...799...37L}
{Leggett}, S.~K., {Morley}, C.~V., {Marley}, M.~S., \& {Saumon}, D. 2015, \apj,
  799, 37, \dodoi{10.1088/0004-637X/799/1/37}

\bibitem[{{Leggett} {et~al.}(2013){Leggett}, {Morley}, {Marley}, {Saumon},
  {Fortney}, \& {Visscher}}]{2013ApJ...763..130L}
{Leggett}, S.~K., {Morley}, C.~V., {Marley}, M.~S., {et~al.} 2013, \apj, 763,
  130, \dodoi{10.1088/0004-637X/763/2/130}

\bibitem[{{Leggett} {et~al.}(2017){Leggett}, {Tremblin}, {Esplin}, {Luhman}, \&
  {Morley}}]{2017ApJ...842..118L}
{Leggett}, S.~K., {Tremblin}, P., {Esplin}, T.~L., {Luhman}, K.~L., \&
  {Morley}, C.~V. 2017, \apj, 842, 118, \dodoi{10.3847/1538-4357/aa6fb5}

\bibitem[{{Liu} {et~al.}(2011){Liu}, {Delorme}, {Dupuy}, {Bowler}, {Albert},
  {Artigau}, {Reyl{\'e}}, {Forveille}, \& {Delfosse}}]{2011ApJ...740..108L}
{Liu}, M.~C., {Delorme}, P., {Dupuy}, T.~J., {et~al.} 2011, \apj, 740, 108,
  \dodoi{10.1088/0004-637X/740/2/108}

\bibitem[{{Lodders} \& {Fegley}(2002)}]{2002Icar..155..393L}
{Lodders}, K., \& {Fegley}, B. 2002, Icarus, 155, 393,
  \dodoi{10.1006/icar.2001.6740}

\bibitem[{{Lodieu} {et~al.}(2013){Lodieu}, {B{\'e}jar}, \&
  {Rebolo}}]{2013AA...550L...2L}
{Lodieu}, N., {B{\'e}jar}, V.~J.~S., \& {Rebolo}, R. 2013, \aap, 550, L2,
  \dodoi{10.1051/0004-6361/201220696}

\bibitem[{{Luhman}(2014)}]{2014ApJ...786L..18L}
{Luhman}, K.~L. 2014, \apjl, 786, L18, \dodoi{10.1088/2041-8205/786/2/L18}

\bibitem[{{Luhman} {et~al.}(2011){Luhman}, {Burgasser}, \&
  {Bochanski}}]{2011ApJ...730L...9L}
{Luhman}, K.~L., {Burgasser}, A.~J., \& {Bochanski}, J.~J. 2011, \apjl, 730,
  L9, \dodoi{10.1088/2041-8205/730/1/L9}

\bibitem[{{Luhman} \& {Esplin}(2016)}]{2016AJ....152...78L}
{Luhman}, K.~L., \& {Esplin}, T.~L. 2016, \aj, 152, 78,
  \dodoi{10.3847/0004-6256/152/3/78}

\bibitem[{{Mamajek} {et~al.}(2015){Mamajek}, {Prsa}, {Torres}, {Harmanec},
  {Asplund}, {Bennett}, {Capitaine}, {Christensen-Dalsgaard}, {Depagne},
  {Folkner}, {Haberreiter}, {Hekker}, {Hilton}, {Kostov}, {Kurtz}, {Laskar},
  {Mason}, {Milone}, {Montgomery}, {Richards}, {Schou}, \&
  {Stewart}}]{2015arXiv151007674M}
{Mamajek}, E.~E., {Prsa}, A., {Torres}, G., {et~al.} 2015, ArXiv e-prints,
  arXiv:1510.07674.
\newblock \doarXiv{1510.07674}

\bibitem[{Marley \& Saumon(2020)}]{marley_mark_2020_3733843}
Marley, M., \& Saumon, D. 2020, {Sonora 2018: Cloud-free, solar C/O substellar
  evolution and photometry},  Zenodo, \dodoi{10.5281/zenodo.3733843}.
\newblock \url{https://doi.org/10.5281/zenodo.3733843}

\bibitem[{Marley {et~al.}(2018)Marley, Saumon, Morley, \&
  Fortney}]{marley_mark_2018_1309035}
Marley, M., Saumon, D., Morley, C., \& Fortney, J. 2018, {Sonora 2018:
  Cloud-free, solar composition, solar C/O substellar atmosphere models and
  spectra}, nc\_m+0.0\_co1.0\_v1.0,  Zenodo, \dodoi{10.5281/zenodo.1309035}.
\newblock \url{https://doi.org/10.5281/zenodo.1309035}

\bibitem[{{Marley} \& {Leggett}(2009)}]{2009and..book..101M}
{Marley}, M.~S., \& {Leggett}, S.~K. 2009, {The Future of Ultracool Dwarf
  Science with JWST}, ed. {Thronson, H.~A., Stiavelli, M., \& Tielens, A.}
  ({Dordrecht: Springer}), 101

\bibitem[{{Marley} {et~al.}(2010){Marley}, {Saumon}, \&
  {Goldblatt}}]{2010ApJ...723L.117M}
{Marley}, M.~S., {Saumon}, D., \& {Goldblatt}, C. 2010, \apjl, 723, L117,
  \dodoi{10.1088/2041-8205/723/1/L117}

\bibitem[{{Marocco} {et~al.}(2019){Marocco}, {Caselden}, {Meisner},
  {Kirkpatrick}, {Wright}, {Faherty}, {Gelino}, {Eisenhardt}, {Fowler},
  {Cushing}, {Cutri}, {Garcia}, {Jarrett}, {Koontz}, {Mainzer}, {Marchese},
  {Mobasher}, {Schlegel}, {Stern}, \& {Teplitz}}]{2019ApJ...881...17M}
{Marocco}, F., {Caselden}, D., {Meisner}, A.~M., {et~al.} 2019, \apj, 881, 17,
  \dodoi{10.3847/1538-4357/ab2bf0}

\bibitem[{{Marocco} {et~al.}(2020){Marocco}, {Kirkpatrick}, {Meisner},
  {Caselden}, {Eisenhardt}, {Cushing}, {Faherty}, {Gelino}, \&
  {Wright}}]{2020ApJ...888L..19M}
{Marocco}, F., {Kirkpatrick}, J.~D., {Meisner}, A.~M., {et~al.} 2020, \apjl,
  888, L19, \dodoi{10.3847/2041-8213/ab6201}

\bibitem[{{Marocco} {et~al.}(2021){Marocco}, {Eisenhardt}, {Fowler},
  {Kirkpatrick}, {Meisner}, {Schlafly}, {Stanford}, {Garcia}, {Caselden},
  {Cushing}, {Cutri}, {Faherty}, {Gelino}, {Gonzalez}, {Jarrett}, {Koontz},
  {Mainzer}, {Marchese}, {Mobasher}, {Schlegel}, {Stern}, {Teplitz}, \&
  {Wright}}]{2021ApJS..253....8M}
{Marocco}, F., {Eisenhardt}, P. R.~M., {Fowler}, J.~W., {et~al.} 2021, \apjs,
  253, 8, \dodoi{10.3847/1538-4365/abd805}

\bibitem[{{Meisner} {et~al.}(2020){Meisner}, {Caselden}, {Kirkpatrick},
  {Marocco}, {Gelino}, {Cushing}, {Eisenhardt}, {Wright}, {Faherty}, {Koontz},
  {Marchese}, {Khalil}, {Fowler}, \& {Schlafly}}]{2020ApJ...889...74M}
{Meisner}, A.~M., {Caselden}, D., {Kirkpatrick}, J.~D., {et~al.} 2020, \apj,
  889, 74, \dodoi{10.3847/1538-4357/ab6215}

\bibitem[{{Miles} {et~al.}(2020){Miles}, {Skemer}, {Morley}, {Marley},
  {Fortney}, {Allers}, {Faherty}, {Geballe}, {Visscher}, {Schneider}, {Lupu},
  {Freedman}, \& {Bjoraker}}]{2020AJ....160...63M}
{Miles}, B.~E., {Skemer}, A. J.~I., {Morley}, C.~V., {et~al.} 2020, \aj, 160,
  63, \dodoi{10.3847/1538-3881/ab9114}

\bibitem[{{Morley} {et~al.}(2014){Morley}, {Marley}, {Fortney}, {Lupu},
  {Saumon}, {Greene}, \& {Lodders}}]{2014ApJ...787...78M}
{Morley}, C.~V., {Marley}, M.~S., {Fortney}, J.~J., {et~al.} 2014, \apj, 787,
  78, \dodoi{10.1088/0004-637X/787/1/78}

\bibitem[{{Morley} {et~al.}(2018){Morley}, {Skemer}, {Allers}, {Marley},
  {Faherty}, {Visscher}, {Beiler}, {Miles}, {Lupu}, {Freedman}, {Fortney},
  {Geballe}, \& {Bjoraker}}]{2018ApJ...858...97M}
{Morley}, C.~V., {Skemer}, A.~J., {Allers}, K.~N., {et~al.} 2018, \apj, 858,
  97, \dodoi{10.3847/1538-4357/aabe8b}

\bibitem[{{Rayner} {et~al.}(2009){Rayner}, {Cushing}, \&
  {Vacca}}]{2009ApJS..185..289R}
{Rayner}, J.~T., {Cushing}, M.~C., \& {Vacca}, W.~D. 2009, \apjs, 185, 289.
\newblock \doarXiv{0909.0818}

\bibitem[{{Reach} {et~al.}(2005){Reach}, {Megeath}, {Cohen}, {Hora}, {Carey},
  {Surace}, {Willner}, {Barmby}, {Wilson}, {Glaccum}, {Lowrance}, {Marengo}, \&
  {Fazio}}]{2005PASP..117..978R}
{Reach}, W.~T., {Megeath}, S.~T., {Cohen}, M., {et~al.} 2005, \pasp, 117, 978,
  \dodoi{10.1086/432670}

\bibitem[{{Roellig} {et~al.}(2004){Roellig}, {Van Cleve}, {Sloan}, {Wilson},
  {Saumon}, {Leggett}, {Marley}, {Cushing}, {Kirkpatrick}, {Mainzer}, \&
  {Houck}}]{2004ApJS..154..418R}
{Roellig}, T.~L., {Van Cleve}, J.~E., {Sloan}, G.~C., {et~al.} 2004, \apjs,
  154, 418, \dodoi{10.1086/421978}

\bibitem[{{Saumon} {et~al.}(2000){Saumon}, {Geballe}, {Leggett}, {Marley},
  {Freedman}, {Lodders}, {Fegley}, \& {Sengupta}}]{2000ApJ...541..374S}
{Saumon}, D., {Geballe}, T.~R., {Leggett}, S.~K., {et~al.} 2000, \apj, 541,
  374, \dodoi{10.1086/309410}

\bibitem[{{Saumon} {et~al.}(2012){Saumon}, {Marley}, {Abel}, {Frommhold}, \&
  {Freedman}}]{2012ApJ...750...74S}
{Saumon}, D., {Marley}, M.~S., {Abel}, M., {Frommhold}, L., \& {Freedman},
  R.~S. 2012, \apj, 750, 74, \dodoi{10.1088/0004-637X/750/1/74}

\bibitem[{{Saumon} {et~al.}(2006){Saumon}, {Marley}, {Cushing}, {Leggett},
  {Roellig}, {Lodders}, \& {Freedman}}]{2006ApJ...647..552S}
{Saumon}, D., {Marley}, M.~S., {Cushing}, M.~C., {et~al.} 2006, \apj, 647, 552,
  \dodoi{10.1086/505419}

\bibitem[{{Saumon} {et~al.}(2003){Saumon}, {Marley}, {Lodders}, \&
  {Freedman}}]{2003IAUS..211..345S}
{Saumon}, D., {Marley}, M.~S., {Lodders}, K., \& {Freedman}, R.~S. 2003, in IAU
  Symposium, Vol. 211, Brown Dwarfs, ed. {E.~Mart{\'{\i}}n}, 345

\bibitem[{{Schneider} {et~al.}(2016){Schneider}, {Cushing}, {Kirkpatrick}, \&
  {Gelino}}]{2016ApJ...823L..35S}
{Schneider}, A.~C., {Cushing}, M.~C., {Kirkpatrick}, J.~D., \& {Gelino}, C.~R.
  2016, \apjl, 823, L35, \dodoi{10.3847/2041-8205/823/2/L35}

\bibitem[{{Schneider} {et~al.}(2015){Schneider}, {Cushing}, {Kirkpatrick},
  {Gelino}, {Mace}, {Wright}, {Eisenhardt}, {Skrutskie}, {Griffith}, \&
  {Marsh}}]{2015ApJ...804...92S}
{Schneider}, A.~C., {Cushing}, M.~C., {Kirkpatrick}, J.~D., {et~al.} 2015,
  \apj, 804, 92, \dodoi{10.1088/0004-637X/804/2/92}

\bibitem[{{Stephens} {et~al.}(2009){Stephens}, {Leggett}, {Cushing}, {Marley},
  {Saumon}, {Geballe}, {Golimowski}, {Fan}, \& {Noll}}]{2009ApJ...702..154S}
{Stephens}, D.~C., {Leggett}, S.~K., {Cushing}, M.~C., {et~al.} 2009, \apj,
  702, 154.
\newblock \doarXiv{0906.2991}

\bibitem[{{Tinney} {et~al.}(2014){Tinney}, {Faherty}, {Kirkpatrick}, {Cushing},
  {Morley}, \& {Wright}}]{2014ApJ...796...39T}
{Tinney}, C.~G., {Faherty}, J.~K., {Kirkpatrick}, J.~D., {et~al.} 2014, \apj,
  796, 39, \dodoi{10.1088/0004-637X/796/1/39}

\bibitem[{{Tokunaga} \& {Vacca}(2005)}]{2005PASP..117..421T}
{Tokunaga}, A.~T., \& {Vacca}, W.~D. 2005, \pasp, 117, 421,
  \dodoi{10.1086/429382}

\bibitem[{{Wright} {et~al.}(2010){Wright}, {Eisenhardt}, {Mainzer}, {Ressler},
  {Cutri}, {Jarrett}, {Kirkpatrick}, {Padgett}, {McMillan}, {Skrutskie},
  {Stanford}, {Cohen}, {Walker}, {Mather}, {Leisawitz}, {Gautier}, {McLean},
  {Benford}, {Lonsdale}, {Blain}, {Mendez}, {Irace}, {Duval}, {Liu}, {Royer},
  {Heinrichsen}, {Howard}, {Shannon}, {Kendall}, {Walsh}, {Larsen}, {Cardon},
  {Schick}, {Schwalm}, {Abid}, {Fabinsky}, {Naes}, \&
  {Tsai}}]{2010AJ....140.1868W}
{Wright}, E.~L., {Eisenhardt}, P.~R.~M., {Mainzer}, A.~K., {et~al.} 2010, \aj,
  140, 1868, \dodoi{10.1088/0004-6256/140/6/1868}

\bibitem[{{Yurchenko} {et~al.}(2011){Yurchenko}, {Barber}, \&
  {Tennyson}}]{2011MNRAS.413.1828Y}
{Yurchenko}, S.~N., {Barber}, R.~J., \& {Tennyson}, J. 2011, \mnras, 413, 1828,
  \dodoi{10.1111/j.1365-2966.2011.18261.x}

\bibitem[{{Yurchenko} \& {Tennyson}(2014)}]{2014MNRAS.440.1649Y}
{Yurchenko}, S.~N., \& {Tennyson}, J. 2014, \mnras, 440, 1649,
  \dodoi{10.1093/mnras/stu326}

\bibitem[{{Zalesky} {et~al.}(2019){Zalesky}, {Line}, {Schneider}, \&
  {Patience}}]{2019ApJ...877...24Z}
{Zalesky}, J.~A., {Line}, M.~R., {Schneider}, A.~C., \& {Patience}, J. 2019,
  \apj, 877, 24, \dodoi{10.3847/1538-4357/ab16db}

\end{thebibliography}

\end{document}